\documentclass[nofootinbib,aps,preprint,pre,showpacs,amssymb,superscriptaddress]{revtex4}

\usepackage{graphicx}
\usepackage{bm}% bold math
\usepackage{epsfig,amsmath, amsfonts, amssymb, xcolor,hyperref,wasysym}
\usepackage{stmaryrd}
\usepackage{soul}
\usepackage{textcomp}
\def\be{\begin{equation}}
\def\ee{\end{equation}}
\def\la{\left<}
\def\ra{\right>}

\def\bea{\begin{eqnarray}}
\def\eea{\end{eqnarray}}

\begin{document}
\title{Action at a distance in classical uniaxial ferromagnetic arrays}

\author{D. B. Abraham}
\affiliation{Theoretical Physics, Department of Physics, University of Oxford,
1 Keble Road, Oxford OX1 3NP, United Kingdom}
\author{A.~Macio\l ek}

\affiliation{Institute of Physical Chemistry,
             Polish Academy of Sciences, Kasprzaka 44/52,
            PL-01-224 Warsaw, Poland}
\affiliation{Max-Planck-Institut f{\"u}r Intelligente Systeme, Heisenbergstr.~3,
D-70569 Stuttgart, Germany}
\author{A. Squarcini}
\affiliation{Max-Planck-Institut f{\"u}r Intelligente Systeme, Heisenbergstr.~3,
D-70569 Stuttgart, Germany}
\affiliation{IV Institut f\"ur Theoretische  Physik,
  Universit\"at Stuttgart, Pfaffenwaldring 57, D-70569 Stuttgart, Germany }

\author{O. Vasilyev}
\affiliation{Max-Planck-Institut f{\"u}r Intelligente Systeme, Heisenbergstr.~3,
D-70569 Stuttgart, Germany}
\affiliation{IV Institut f\"ur Theoretische  Physik,
  Universit\"at Stuttgart, Pfaffenwaldring 57, D-70569 Stuttgart, Germany }

\date{\today}

\begin{abstract}
We examine in detail the theoretical foundations of  striking long-range couplings emerging in  arrays of fluid cells  connected by narrow channels by using a lattice gas (Ising model) description of a system. We present a reexamination of the well known exact determination of the two-point correlation function along the edge of a channel using the transfer matrix technique and a new interpretation is provided. The explicit form of the correlation length is found to grow exponentially with the cross section of the channels at the bulk two-phase coexistence. The aforementioned result is recaptured by a refined version of the Fisher-Privman theory of first order phase transitions in which the Boltzmann factor for a domain wall is decorated with a contribution stemming from the point tension originated at its endpoints. The Boltzmann factor for a domain wall together with the point tension is then identified exactly thanks to two independent analytical techniques, providing a critical test of the Fisher-Privman theory. We then illustrate how to build up the network model from its elementary constituents, the cells and the channels. Moreover, we are able to extract the strength of the coupling between cells and express them in terms of the length and  width and coarse grained quantities such as surface and point tensions. We then support our theoretical investigation with a series of corroborating results based on Monte Carlo simulations. We illustrate how the  long range ordering occurs and how the latter is signaled by the thermodynamic quantities corresponding to both planar and three-dimensional Ising arrays.
\end{abstract}


\maketitle

%====================================
%================= SECTION 1
%====================================
\section{Introduction}
\label{sec:int}
\begin{figure}[htbp]
\includegraphics[width=80mm]{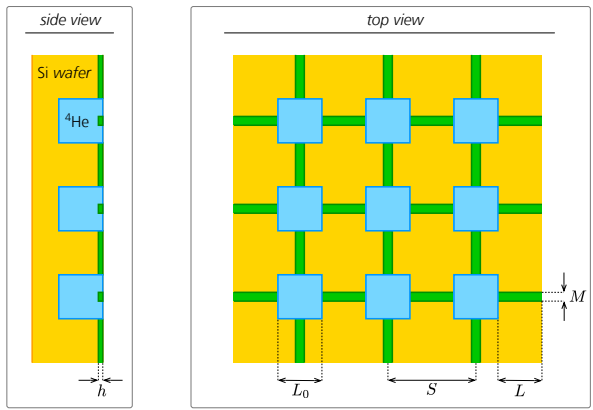}
\centering
\caption{The network of boxes from top view (right) and from side view (left). The typical dimensions considered in the experimental setup (see text) are: $L \sim 2 \, \mu\textrm{m}$, $L_{0} \sim 2 \, \mu\textrm{m}$, $h \sim 30 \, \textrm{nm}$.}
\label{fig_array}
\end{figure}

Recent experimental work by Gasparini et al. \cite{PKMG} has demonstrated striking action-at-a-distance effects in superfluid ${}^{4}$He. The typical system is formed from a two-dimensional array of identical microscopic boxes etched in a Si wafer. These are filled with liquid ${}^{4}$He and then coupled by pouring a relatively thin supernatant layer of liquid ${}^{4}$He on top. Another technique for achieving coupling is to use a network of fluid channels \cite{PG}. The signature of action-at-a-distance is provided by accurate thermodynamic measurements, which show a quite unexpected ``shoulder''. The dimensions used in these experiments can be appreciated from Fig.\ref{fig_array}.

A crucial factor in this phenomenology is the proximity induced by both the size and connectivity of the boxes, together with the nearness to the critical point. A precise discussion on the relevance of proximity effects on the enhancement of ordering in the contest of Gasparini et al. experiment can be found in \cite{Fisher}. Perron et al. \cite{PG1} suggested that this class of experimental results might be a more widespread consequence of the critical phenomenon than previous supposed. Stimulated by these remarks, we have shown \cite{AMV}  that for Ising systems (an entirely different universality class), there is a divergent length scale (not the usual critical one) which is responsible for emergent long-ranged effects. This brings together ideas of Kac \cite{Kac} on asymptotic spectral degeneracy in transfer matrices, the Fisher-Privman \cite{privman_fisher} theory of finite size effects in first order phase transitions (and other systems) and the appropriate solution for the planar Ising model on a strip with free boundary conditions \cite{DBA:71}. The latter shows first how effective the Fisher-Privman theory is when accurate input data are used. To provide this accurate input data, we give another exact solution, which gives as a bonus an exact result for the point tension but in a different context from \cite{point_tension}. Our thinking is illustrated and extended by some Monte-Carlo simulations.

The layout of the paper is as follows: in Sec. \ref{subsec:cor} we summarize the calculation of the pair correlation function for spins located in the edges of the strip \cite{DBA:71}. The algebra of the original derivation \cite{DBA:71} is quite heavy, so we have focused here on the physical motivation, drawing analogies with the quantum mechanics of spinless fermions on a finite line (with ends, rather than the more transparent case with cyclic boundary conditions). Then in Sec. \ref{subsec:FP}, the same problem is treated in a completely different way using the Fisher-Privman theory \cite{privman_fisher}. We point out that introducing a hypothetical point tension in the statistical weight of an isolated domain wall gives precise agreement with part of the exact solution in Sec. \ref{subsec:cor}. Normally, such contributions appear to be ignored but they are mandatory if critical scaling is to be captured. We also consider how an effective coupling is to be set up in a ``network'' lattice of boxes (each 
characterised by an up/down magnetisation, which will be valid for large enough boxes). These ``boxes'' are coupled by strips in which the internal degrees of freedom have been summed out, producing an Ising superlattice of nodes which can display long range order which, since the effective coupling is temperature-dependent, is far from obvious.

After this, in Sec. \ref{subsec:dw} we derive the Fisher-Privman weight from first principles in an Ising strip and show that it has exactly the value deduced on phenomenological grounds. The fact that this can be done shows how good the Fisher-Privman theory is when an appropriate weight is used. Our final remarks before reporting the results of simulations will be discussions of the Kac theory of asymptotic degeneracy in transfer matrix spectra (Sec. \ref{subsec:asymptotic_degeneracy}).

%====================================
%================= SECTION 2
%====================================
\section{Theory}
\label{sec:theory}
%====================================
%================= SECTION 2.A
%====================================
\subsection{Correlation function}
\label{subsec:cor}
\begin{figure}[htbp]
\includegraphics[width=120mm]{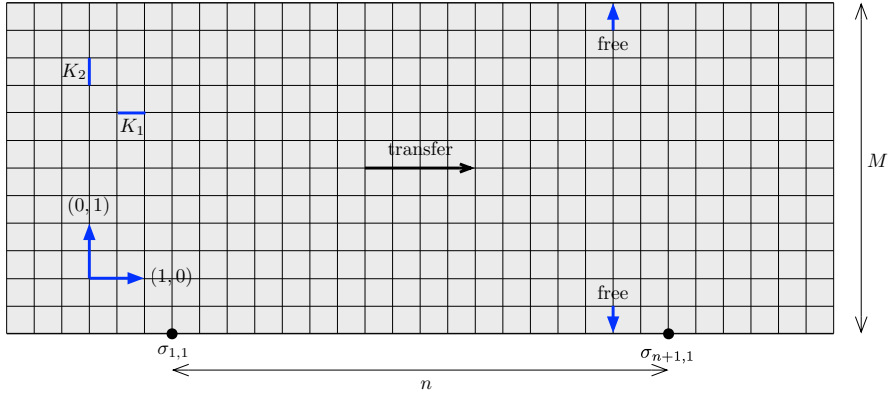}
\centering
\caption{Ising model on a planar lattice with free boundary conditions and nearest-neighbour interactions $K_j>0, j=1, 2$ is shown. The transfer direction is indicated. In the applications considered here, we impose cyclic boundary conditions in the $(1,0)$ direction.}
\label{fig:1}
\end{figure}

We consider the correlation function between two spins in the edge of a planar Ising Ferro-magnet with zero magnetic field and strip geometry  as shown in Fig.~\ref{fig:1}. The significance of having the spins in an edge is that this calculation is particularly tractable in the transfer matrix language \cite{DBA:71} and its correct interpretation leads to a significant enhancement of the Fisher-Privman theory of finite-size effects \cite{privman_fisher}. The transfer matrix calculation builds up the lattice from column  to column (see Fig.~\ref{fig:1}) with the operator:
\begin{equation}
 \label{eq:t1}
 V_1 = (2\sinh 2K_{1})^{M/2}\exp\left[ -K_1^*\sum_{m=1}^M\sigma_m^z\right] \, ,
\end{equation}
where $\tanh K_{1}=\textrm{e}^{-2K_{1}^{*}}$. Here we are using the Schultz, Mattis and Lieb \cite{SML} convention for spin operators $\sigma_m^i, i=x, y, z$ with {\it ordered} direction taken as $x$. The  pre-factor of  $(2\sinh 2K_1)^{M/2}$ will always cancel out with the normalizing partition function and thus we omit it for simplicity, as it will not report in the final answer. The transfer matrix $V_2$  which accounts for the interactions within columns is of the diagonal form
\begin{equation}
 \label{eq:t2}
 V_2 = \exp\left[K_2\sum_{m=1}^{M-1}\sigma_m^x\sigma_{m+1}^x\right]
\end{equation}
for strip boundary conditions as in Fig.~\ref{fig:1}. The spectrum of the symmetrized product $V=V_2^{1/2}V_1V_2^{1/2}$ was determined some time ago \cite{DBA:71} by an amalgamation of the techniques of Kaufman \cite{kaufmann} and of Schultz, Mattis and Lieb (SML) \cite{SML}. It was Kaufman who made the essential step of introducing the Jordan-Wigner transformation \cite{JW}, which reduces the rather intractable spin problem to one involving quadratic forms of spinors. Her method of diagonalisation was made more tractable by SML, who drew an analogy with the pairing ideas of Anderson \cite{anderson} and of Nambu \cite{nambu}. Essentially, if one has a good working knowledge of the Bardeen, Cooper and Schrieffer theory of superconductivity \cite{BCS}, then the Onsager theory \cite{O44} has been brought within the pabulum of any reasonably well-educated theoretical physicist.

The expression for the edge-spin pair correlation function is:
\begin{equation}
\label{eq:t3}
C(n)=\langle\sigma_{1,1}\sigma_{1,n+1}\rangle = \langle\Phi|\sigma_1^xV^n\sigma_1^x|\Phi\rangle\Lambda_0^{-n} \, ,
\end{equation}
where $|\Phi\rangle$ is the maximal eigenvector (unique by the Perron-Frobenius theorem \cite{Lax}) with eigenvalue $\Lambda_0$.This result has been obtained by imposing periodic boundary conditions in the strip axial direction and then taking the limit of infinite strip length.

The Jordan-Wigner transformation, which is the crucial step introduced by Kaufman, as we stated above, is given by
\begin{equation}
\label{eq:t4}
 f_m^{\dagger} = P_{m-1} \left( \sigma_m^x+i\sigma_m^y \right)/2 , \quad  2 \leqslant m \leqslant M \quad \mathrm{and} \quad   f_1^{\dagger} = \left( \sigma_1^x+i\sigma_1^y \right)/2 \, ,
\end{equation}
with the Jordan-Wigner tail or string being specified by
\begin{equation}
 \label{eq:t5}
 P_m=\prod_{j=1}^m\left(-\sigma_j^z\right)=i^m\exp\left[-i\sum_{j=1}^m\sigma_j^z\right] \, .
\end{equation}
The Fermi field obey the anti-commutation relations $\left[f_m,f_n^{\dagger}\right]_+=\delta_{mn}, \left[f_m,f_n\right]_+=0$. The connection of the operator $P_M$ with spin rotations (inversions) in a column of a lattice is now obvious, as are the commutation relations
\begin{equation}
\label{eq:t6}
\left[V_1, P_M\right] =  \left[V_2, P_M\right] = 0 \, .
\end{equation}
This implies that we can seek simultaneous eigenvectors of $V$ and $P_M$. The $V_j, j=1,2$ become quadratic forms in fermions:
\begin{equation}
\label{eq:t7}
V_1=\exp\left[-K_1^*\sum_{m=1}^M\left(2f_m^{\dagger}f_m-1\right)\right] \, , \qquad V_2=\exp\left[K_2\sum_{m=1}^{M-1}(f_m^{\dagger}-f_m)(f_{m+1}^{\dagger}+f_{m+1} )\right] \, .
\end{equation}
The correlation function in (\ref{eq:t3}) is expressed in terms of the Fermi fields as
\begin{equation}
\label{eq:t8}
C(n)=\langle\Phi|\left(f_1^{\dagger}+f_1\right)V^n\left(f_1^{\dagger}+f_1\right)|\Phi\rangle\Lambda_0^{-n} \, .
\end{equation}
A consequence of taking the spins in the edge is the {\it linearity} of the forms representing the quantum mechanical treatment of spinless fermions on a finite line. We have hopping terms, which correspond to kinetic energy, and on-site energy terms. Thus we anticipate left and right going waves characteristic of the bulk, which may be compared with the SML solution. The ``in'' and ``out'' waves at the left boundary must be matched to fit the boundary conditions; they become ``out'' and ``in'' waves at the right boundary, which must also be matched and made compatible with the left boundary. This generates the  discretization condition for the fermion momentum. The only additional feature is that on the lattice there may be a local modification of amplitudes at the ends of the column. We find that
\begin{equation}
\label{eq:t9}
V=\exp\left[-\sum_{k}\gamma(k)\left(X^{\dagger}(k)X(k)-1/2\right)\right] \, ,
\end{equation}
where $X(k)$ the are Fermi operators, details of which will follow and $\gamma(k)$, the celebrated Onsager function \cite{O44}, is that solution of
\begin{equation}
\label{eq:t10}
\cosh \gamma(k) = \cosh 2K_1^*\cosh 2K_2 - \sinh 2K_1^*\sinh 2K_2\cos k \, ,
\end{equation}
which is non-negative for real $k$. This requirement makes the vacuum for the operators also the maximal eigenvector: $X(k)|\Phi\rangle =0$.

The discretisation condition mentioned above for $k$ is
\begin{equation}
\label{eq:t11}
\textrm{e}^{iMk} = s \, \textrm{e}^{i\delta^*(k)}, \qquad s=\pm 1 \, ,
\end{equation}
where $s$ encodes reflection behaviour \cite{AM,AM_1} of the eigenvectors and the angle $\delta^*(k)$, also introduced by Onsager, is defined by
\begin{equation}
\label{eq:t12}
\textrm{e}^{i\delta^*(k)} = \left(\frac{B}{A}\right)^{1/2}\left[\frac{\left(\textrm{e}^{ik}-A\right)\left(\textrm{e}^{ik}-B^{-1}\right)}{\left(\textrm{e}^{ik}-A^{-1}\right)\left(\textrm{e}^{ik}-B\right)}\right]^{1/2} \, .
\end{equation}
The location of the square-root branch points in the above is determined by
\begin{equation}
\label{eq:t13}
A= \exp\left(K_1+K_2^*\right) \quad\quad \mathrm{and} \quad \quad B = \exp\left(K_1-K_2^*\right) \, .
\end{equation}
The choice of the branch for $\gamma(k)$  determines that in the discretisation equation (\ref{eq:t12}). For sub-critical temperatures, we have  $K_1 > K_2^* >0$ ; thus $A>B>1$  and $\delta^*(0) = 0$ (mod $2\pi$). It is convenient to define, as did Kaufman, the spinors by:
\begin{equation}
\label{eq:t14}
\Gamma_{2m-1} = f^{\dagger}_m + f_m, \quad \quad  \Gamma_{2m} = -i\left(f^{\dagger}_m - f_m\right),  \quad \quad m=1,\ldots, M \, .
\end{equation}
These have a simple representation in terms of the $X(k)$ which is useful for calculating correlation functions; this is
\begin{equation}
\label{eq:t15}
\Gamma_m = \sum_kN(k)\left[y^*_m(k)X^{\dagger}(k)+y_m(k)X(k)\right] \, ,
\end{equation}
where the normalisation factor $N(k)$ is not needed in the computation and
\begin{equation}
\label{eq:t16}
y_{2m}(k) = iN(k)\sin mk, \quad y_{2m+1}(k) = N(k)\sin\left(km-\delta^*(k)\right), \quad m=1,\ldots, M-1 \, ,
\end{equation}
with the boundary values:
\begin{equation}
\label{eq:t17}
y_1(k) = -N(k)\cosh K_2\sin\delta^*(k), \quad y_{2M} = -isN(k)\cosh K_2\sin\delta^*(k) \, .
\end{equation}
Here we see the intuitive ideas above, between (\ref{eq:t8}) and (\ref{eq:t9}), in action. Taken with the discretization condition, (\ref{eq:t15}), (\ref{eq:t16}) and (\ref{eq:t17}) guarantee 
\begin{equation}
\label{eq:t18}
\left[X(k_1), X(k_2)\right] = 0, \qquad \left[X(k_1), X^{\dagger}(k_2)\right] = \delta_{k_1,k_2} \, .
\end{equation}
Notice that $k=0$ and $k=\pi$ generate trivial solutions and that, to avoid repetition and triviality, the momenta should satisfy $0 < k < \pi$. In order to calculate the edge pair correlation function, the first step is to determine the allowed  momenta. This is an elementary matter using techniques from elementary calculus. If we consider zeros of $Mk-j\pi-\delta^*(k)$ for $A > B > 1$, there is one (in fact at least one) for each $j = 1,\ldots, M-1$. With $j=0$, there is a non-trivial one if $M < \textrm{d}\delta^{*}(\omega)/\textrm{d}\omega \vert_{\omega=0}$ the slope of $\delta^*$ at $k=0$ and when $M > \textrm{d}\delta^{*}(\omega)/\textrm{d}\omega \vert_{\omega=0}$, there is one with a pure imaginary wavenumber with $s=+1$. Thus $k=iv$, $v$ real and
\begin{equation}
\label{eq:t19}
\textrm{e}^{-Mv} = s \, \textrm{e}^{i\delta^*(iv)} \, .
\end{equation}
It is then a straightforward matter to show there is such a solution for $s=+1$, $0<v<\hat\gamma(0), \hat\gamma(0) = 2(K_1-K_2^*)$, but only if $M > \textrm{d}\delta^{*}(\omega)/\textrm{d}\omega\vert_{\omega=0}$; note that $\hat\gamma(k)$ is just $\gamma(k)$ with $K_1$ and $K_2$ interchanged. It is easy to see that
\begin{equation}
\label{eq:t20}
v=\hat\gamma(0)-2(\sinh2K_1\sinh2K_2)^{-1}\sinh\hat\gamma(0) \, \textrm{e}^{-2M\hat\gamma(0)}+\mathcal{O}(\textrm{e}^{-4M\hat\gamma(0)}) \, ,
\end{equation}
and that 
\begin{equation}
\label{eq:t21}
\gamma(iv) = 2\sinh2K_{1}^{*}\sinh\hat\gamma(0) \, \textrm{e}^{-M\hat\gamma(0)}+\mathcal{O}(\textrm{e}^{-2M\hat\gamma(0)}) \, .
\end{equation}
Thus we have found an asymptotic degeneracy in the spectrum (see Sec.~\ref{subsec:asymptotic_degeneracy}) and this is associated with a surface mode in which the eigenfunctions decay away from the surface on a scale of $\xi_b = 1/\hat\gamma(0)$; this is the bulk correlation length, up to the Kadanoff-Wu anomaly \cite{KW}. The edge-spin pair correlation function comes out in the form
\begin{equation}
\label{eq:t22}
C(n)=m_e^2\exp\Bigl[-2n \sinh2K_{1}^{*}\sinh\hat\gamma(0) \, \textrm{e}^{-M\hat\gamma(0)}\Bigr]+\sum_k|y_1(k)|^2\exp\bigl[-n\gamma(k)\bigr] \, .
\end{equation}
The first term above decays to zero on a new, emergent, length scale $\varpropto (\sinh\hat{\gamma}(0))^{-1}\textrm{e}^{M\hat\gamma(0)}$  on which long-ranged order is ultimately lost. It also displays a scaled form in the vicinity of the bulk correlation length, that is the bulk scaling limit. In the above, $m_e$ is the edge spontaneous magnetization, as originally determined by McCoy and Wu \cite{MW}. The second term on the right hand side is bounded above by $\exp(-n\gamma(0))$; this gives a clear separation of length scales.

%====================================
%================= SECTION 2.B
%====================================
\subsection{Fisher-Privman theory applied to the strip}
\label{subsec:FP}
\begin{figure}[htbp]
\includegraphics[width=160mm]{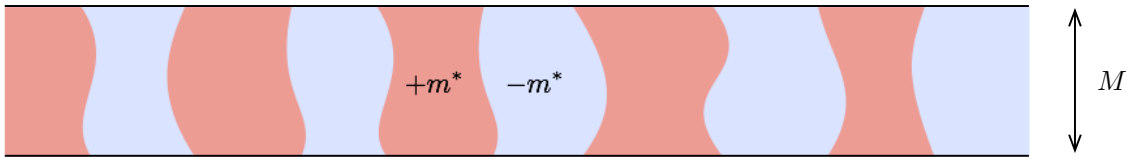}
\centering
\caption{A schematic representation of a typical domain wall configuration on the Ising strip. Typically, the domains are widely separated along $(1,0)$ and thus infrequent.}
\label{fig_FP}
\end{figure}

Any configuration of the spins on the Ising strip with free boundary conditions can be analysed to extract arrangements of domain walls going from side to side of the strip; these walls separate oppositely magnetized domains which are themselves reasonable approximations to bulk strip states for $M$ large enough. We may consider that fluctuation effects with a spatial extent of about the bulk correlation length or less have been summed over, a coarse graining producing a meso-scale model (at least in principle). In this case, the space between domain walls is essentially featureless, having a spatially-averaged magnetisation of bulk spontaneous magnetisation, denoted $m^{*}$; see Fig.\ref{fig_FP}.
The energy of any domain wall should be replaced by a coarse-grained fluctuation free energy of Helmholtz type. Following Fisher and Privman \cite{privman_fisher}, we can go one step further and regard the domain walls as point particles in a quasi one-dimensional system, the equilibrium statistical mechanics of which can be determined exactly in a suitable approximation. Let the statistical weight of any domain wall in isolation be denoted by $\widetilde{w}$. Then the two spins in the strip separated by $n$ lattice spacings will be parallel (resp. anti-parallel) if the number of domain walls in the interviewing space are even (resp. odd). The correlation function of these spins, denoted $\mathcal{C}(n)$, is given by
\begin{eqnarray}\nonumber
\label{sec_FP_01}
\mathcal{C}(n) & = & \biggl\{ \frac{1}{2} \bigl[ (1+\widetilde{w})^{n} + (1-\widetilde{w})^{n} \bigr] - \frac{1}{2} \bigl[ (1+\widetilde{w})^{n} - (1-\widetilde{w})^{n} \bigr] \biggr\} / \bigl(1+\widetilde{w}\bigr)^{n} \, , \\
& = & \bigl[ (1-\widetilde{w}) / (1+\widetilde{w})\bigr]^{n} \, .
\end{eqnarray}
Here, we assume the domain walls have negligible interactions. Evidently, we have
\begin{equation}
\label{sec_FP_02}
\mathcal{C}(n) = \textrm{e}^{-n\lambda(\widetilde{w})} , \qquad \lambda(\widetilde{w}) = \ln\bigl[(1+\widetilde{w})/(1-\widetilde{w})\bigr] = 2 \Bigl[ \widetilde{w} + 3^{-1} \widetilde{w}^{3} + \mathcal{O}\left(\widetilde{w}^{5}\right) \Bigr] \, .
\end{equation}
Thus, for small $\tilde w$, where the theory is likely to be particularly pertinent, we have $\lambda(\widetilde{w}) \simeq 2\widetilde{w}$. The usual practice is to write
\begin{equation}
\label{sec_FP_03}
\widetilde{w}=\textrm{e}^{-M\tau} \, ,
\end{equation}
for a strip of width $M$, where $\tau$ is the surface tension, a coarse-grained entity as we would expect; this is precisely what one would normally do in Helmholtz fluctuation theory. This do not agree with (\ref{eq:t22}), which is the exact solution for the edge pair function. It would agree, were $\widetilde{w}$ to be replaced by $w$, where
\begin{equation}
\label{sec_FP_03.01}
w = \sinh2K_{1}^{*}\sinh\hat{\gamma}(0) \, \textrm{e}^{-M\hat{\gamma}(0)} \, .
\end{equation}
Since $\hat{\gamma}(0)=\tau$, where $\tau$ is the surface tension for an interface oriented at right angles to the strip axis, we recapture (\ref{eq:t22}) from (\ref{sec_FP_02}) above in the linear regime. Another more phenomenological angle is to note that the Fisher-Privman result above does not scale, but it would do so, were we to write
\begin{equation}
\label{sec_FP_03.02}
w = a \, \xi_{b}^{-1} \textrm{e}^{-M\tau} \, ,
\end{equation}
where $\xi_{b} $ is the bulk correlation length, related precisely to $\tau$ by the relation $\tau\xi_{b}=1/2$, which is valid for all temperatures and is an application of duality \cite{AGM,FF}. It is also compatible with Widom scaling \cite{Widom}. As it stands, if all we demanded was scaling rather than agreement with (\ref{eq:t22}), then the parameter $a$ in (\ref{sec_FP_03.02}) would be an arbitrary scale factor. In section \ref{subsec:dw}, we will calculate $w$ by another exact solution for the Ising strip and see that it is precisely of the form of (\ref{sec_FP_03.01}). Also, we should think of the pre-factor, which converts (\ref{sec_FP_03}) to (\ref{sec_FP_03.01}) as arising from point tension contributions of magnitude $\tau_{p}$. In other words we can write $w=\textrm{e}^{-2\tau_{p}-M\tau}$ and single out the factor $a/\xi_{b}$ as the one due to the point tension. The incorporation of the point tension in the Boltzmann weight defines the enhanced Fisher-Privman theory, but it might just as well be said to be Fisher-Privman theory properly carried out.

With $M$ fixed, $w$ may ultimately be reduced by going towards the critical point. This is contrary to the usual intuition about such matters and it enhances the domain of validity of the Fisher-Privman theory in an interesting way, as we shall see later. Finally, we discuss scaling. If we take the scaling limit $M \rightarrow \infty$, $\tau \rightarrow 0$, $M\tau\rightarrow \overline{M}$, $n\tau\rightarrow \overline{n}$, we see that the non-linear terms in (\ref{sec_FP_02}) generate corrections to scaling. For simplicity, take the isotropic lattice with $K_{1}=K_{2}$, so that $\sinh2K_{1}^{*}=1$ at $\tau=0$. Then we have
\begin{equation}
\label{ }
\mathcal{C}(n) = \exp\Bigl[ -2\overline{n} \textrm{e}^{-\overline{M}} \Bigr] + \mathcal{O}\left(\tau^{2}\right) \, .
\end{equation}
Terms of higher order in $\textrm{e}^{-\overline{M}}$ have been neglected in this equation, since for consistency, we would have to consider higher order terms in (\ref{eq:t21}).

\begin{figure}[h!]
\begin{center}
\includegraphics[width = 0.7\textwidth]{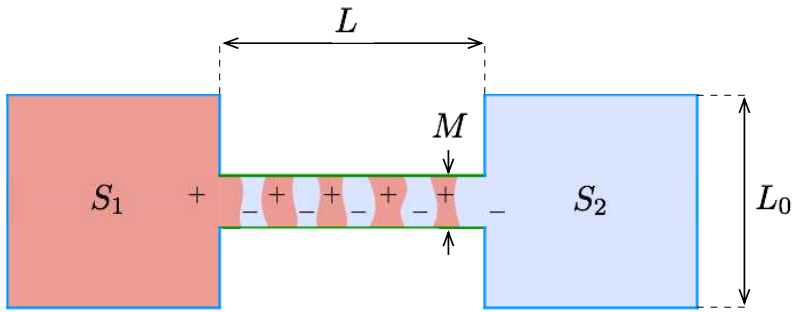}
\caption{Side view  of an Ising system comprised of two cubic lattice boxes of a side $L_0$ connected by a $L \times M$ strip with $L \gg M$. We assume $L_0 \gg M$. 
\label{fig:4}}
\end{center}
\end{figure}

We now consider network models of hyper-cubic coupled ``boxes''. Each one is a finite Ising lattice in $d$-dimensions, with $d=2,3$ and side $L_0$. The interactions in the box and $L_0$ are chosen to ensure that each such box contains essentially a single magnetic domain. Multiple domains, associated with domain walls that intersect the boundary of the box, are suppressed by a strictly positive surface tension (chosen large enough) and the extent of such a domain wall, $\sim L_0^{d-1}$. Thus any such box $j$ has an average magnetization $m_{0}^{*}S_{j}$ where $S_{j}=\pm1$ is an indicator variable and $m_{0}^{*}$ approximates the spontaneous magnetization in $d$-dimensions as $L_0\rightarrow\infty$. In order to investigate coupling within this network let boxes $i$ and $j$ be connected by a strip or rod of Ising type (see Fig.~\ref{fig:4}). 

Then, if $S_{i}S_{j}=+1$ (resp. $-1$), there is an even (resp. odd) number of domain walls within the connector; these are treated by Fisher-Privman theory. The result is to generate a Boltzmann factor $A \, \textrm{e}^{K_{\textrm{eff}}S_{i}S_{j}}$ where the parameter $A$ will be of no further interest, but $\textrm{e}^{2K_{\textrm{eff}}}=\bigl[(1+w)^{L}+(1-w)^{L}\bigr]/\bigl[(1+w)^{L}-(1-w)^{L}\bigr]$, where $L$ is the strip length and $w$ is the \emph{a priori} weight of any domain wall, dependent as we have seen on the strip width $M$ and the surface tension $\tau$, as in (\ref{sec_FP_03.02}). Introducing the variable $t=(1-w)/(1+w)$ this has the form
\begin{equation}
\label{Boltzmann}
\textrm{e}^{2K_{\textrm{eff}}} = \left( 1 + t^{L} \right) / \left( 1 - t^{L} \right) \, ,
\end{equation}
with (see Sec. \ref{subsec:dw} for $K_{1}=K_{2} \equiv K$)
\begin{equation}
\label{weight}
w = (\sinh2K)^{-1}\sinh\tau \, \textrm{e}^{-M\tau} \, .
\end{equation}
It is crucial to note that $K_{\textrm{eff}}$ is an effective Ising coupling which depends in a quite subtle way on $M$, $\tau$ and $K$ (the spin coupling within the strip).

\begin{figure}[htbp]
\includegraphics[width=160mm]{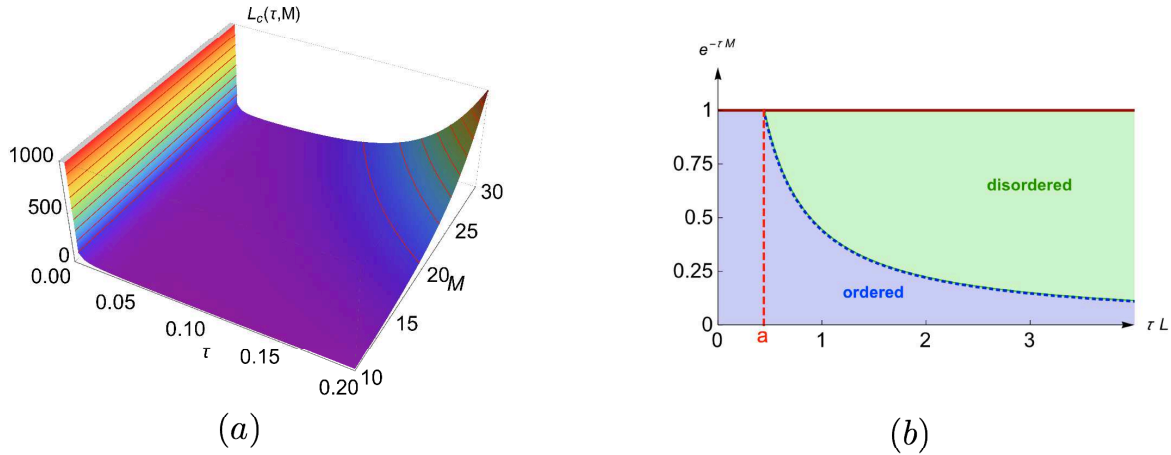}
\caption{($a$) the critical length $L_{c}$ as a function of the surface tension $\tau$ and the strip width $M$. Given $\tau$ and $M$, a system with a length smaller than $L_{c}$ is ordered and corresponds to a point in the phase diagram below the surface of the graph. The iso-$L_{c}$ contour lines are highlighted in red. Note the existence, for a given $M$, of ordered configurations for a pair of values of $\tau$ (reentrant phenomenon). ($b$) the phase diagram in terms of the scaling variables $\tau L$ and $\textrm{e}^{-\tau M}$. The critical line $\tau L_{c}\textrm{e}^{-\tau M}=2^{-1}\ln(1+\sqrt{2})$ separates ordered and disordered configurations, as illustrated in the shadowed regions. Notice that $\textrm{e}^{-\tau M}$ is bounded from above by unity since $\tau$ is non negative.}
\label{fig_phasediagram}
\end{figure}

Let us now consider  the planar array of square boxes connected by one-dimensional Ising rods. Thus it is an interesting question whether the network can display long range order. This would be so if $\textrm{e}^{2K_{\textrm{eff}}}$ can be chosen to exceed the critical value of $1+\sqrt{2}$ of the $d=2$ Ising model. Thus the equation
\begin{equation}
\label{ }
1+t^{L_{c}} = (1+\sqrt{2}) \left( 1-t^{L_{c}} \right)
\end{equation}
implies a critical surface $L_{c}(\tau,M)$, shown in Fig.\ref{fig_phasediagram}($a$). For $L<L_{c}$ (resp. $L>L_{c}$), the network is ordered (resp. disordered). Introducing scaling variables $\tau L_{c}$ and $\textrm{e}^{-M\tau}$ the network critical point satisfies
\begin{equation}
\label{ }
\tau L_{c} \, \textrm{e}^{-M\tau} = 2^{-1}\ln(1+\sqrt{2}) \, .
\end{equation}
Thus, if $\tau L_{c} < 2^{-1}\ln(1+\sqrt{2})$, no such critical point is possible. This is shown in Fig.\ref{fig_phasediagram}($b$).

The role played by the point tension and the domain of validity of the refined Fisher Privman theory can be neatly appreciated with the following considerations. 
Comparing the contribution of the imaginary wavenumber mode in the exact Ising strip calculation, (\ref{eq:t22}), and the result of the refined Fisher Privman model ((\ref{sec_FP_02}) with $\tilde w$ replaced by $w$ given by (\ref{weight})), we should require for perfect matching that:
\begin{equation}
\frac{1-w}{1+w} = \exp\bigl[-\gamma\left(iv(M)\right)\bigr] \, ,
\end{equation}
or $w=\tanh\left(\gamma(iv(M))/2\right)$. Now if we take the exact calculation of the weight, we get: $w=\gamma(iv(M))/2$, which agrees precisely to first order in $\gamma(iv(M))\rightarrow0$. The Fisher-Privman model neglects interactions between domain walls, other than a simple exclusion (walls cannot cross). This cannot be precisely correct. Now examine $\gamma(iv(M))$ given by (\ref{eq:t21}) with $\hat\gamma(0) = \tau$,
\begin{equation}
\gamma(iv(M)) = 2\sinh2K_{1}^{*}\sinh\tau \, \textrm{e}^{-M\tau} + \mathcal{O}\left(\textrm{e}^{-2M\tau}\right) \, ,
\end{equation}
just considering the first term, the behaviour as a function of $\tau$ with $M$ fixed implies investigating the function: $\varphi(\tau) = \sinh\tau \, \textrm{e}^{-M\tau}$. 
Now $\varphi(0)=0$, which is a consequence of the line tension ($\textrm{e}^{-\tau_p}\propto \sinh \tau$). On the other hand for $M\geqslant2$, we have $\lim_{\tau\rightarrow\infty}\varphi(\tau)=0$, hence there is a maximum when $\coth\tau_{m}=M$, or equivalently, $\sinh\tau_{m}=(M^{2}-1)^{-1/2}$. After some algebra we find that the maximum value of $\varphi(\tau)$ is
\begin{equation}
\varphi(\tau_{m}) = \frac{(M-1)^{\frac{M-1}{2}}}{(M+1)^{\frac{M+1}{2}}} = \frac{1}{M\textrm{e}} \left( 1 + \mathcal{O}\left(\frac{1}{M}\right)\right) \, ,
\end{equation}
it is evident that the above can be made as small as one likes by taking $M$ big enough. Consequently the corresponding weight will be \emph{small even for $\tau \rightarrow 0$ thanks to the point tension}; notice that this would not have been the case without the point tension contribution which would have made $w=\mathcal{O}(1)$ in that precise limit. It follows that the refined Fisher-Privman approach is considerably more useful than one might have suspected.

%====================================
%================= SECTION 2.C
%====================================
\subsection{Domain wall free energy}
\label{subsec:dw}
\begin{figure}[htbp]
\includegraphics[scale=0.3]{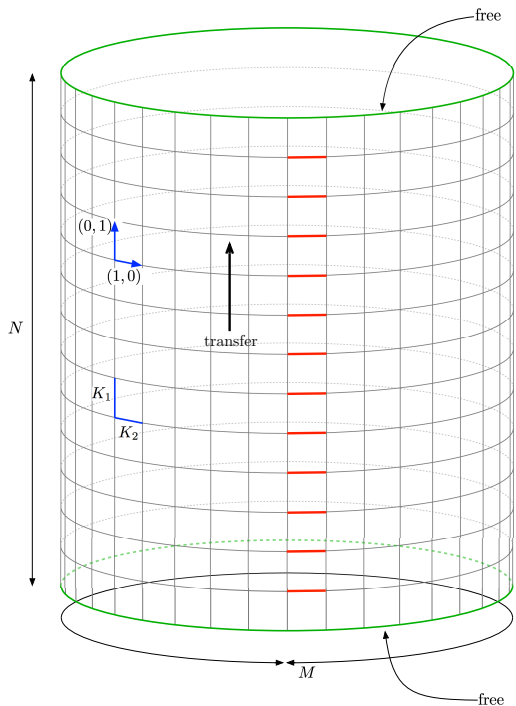}
\caption{Cylindrical lattice with a line of reversed bonds.}
\centering
\label{fig_2}
\end{figure}

The free energy associated with the domain wall running perpendicularly to the  edges of  the strip can be calculated in two different ways. Following earlier definitions, if the strip with free edges (no magnetic fields) is wrapped onto a cylinder, as required by cyclic boundary conditions, then a {\it single} interface in the $(0,1)$ direction can be introduced by reversing a contiguous line  bonds  as shown in the Fig.~\ref{fig_2}.

\begin{figure}[htbp]
\includegraphics[width=7.7cm]{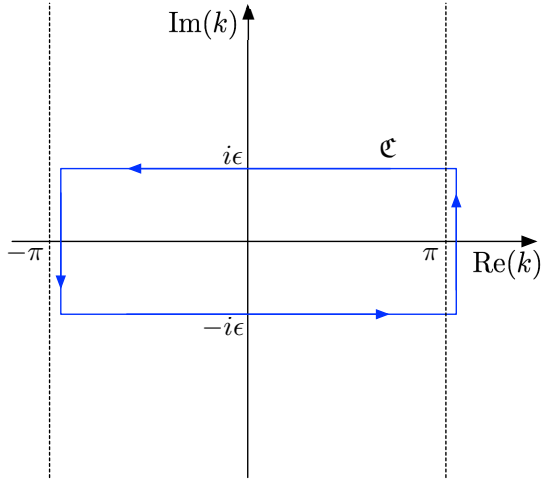}
\centering
\caption{The integration contour $\mathfrak{C}$ in the complex $k$-plane. The vertical lines have $\textrm{Re}(k)=\pi\bigl[1+1/(4M)\bigr]$ and $\textrm{Re}(k)=-\pi\bigl[1-1/(4M)\bigr]$, so that they pass between zeros of $\textrm{e}^{ikM}=\pm1$, and $\pi$ is inside the contour, but not $-\pi$. Note that $g_{N}(k)$ is even in $k$ and there is $2\pi$-periodicity. Thus the side line contributions cancel.}
\label{fig_contour}
\end{figure}

This statement is not quite correct: for sufficiently large $M$ the line of defect bonds admits an {\it odd} number of interfaces, stricto sensu. For temperatures below the critical value for the bulk lattice, an incipient ordered state is expected, so provided the circumference is not too large, in a way that will be made precise in due course, there is a single magnetised phase having average magnetisation approximately the bulk spontaneous value. Introducing a line of reversed bonds as shown in Fig.~\ref{fig_2} will then indeed model an interface. The reader may like to note that this is not unlike the model of an interface from which Onsager extracted the first exact result for the surface tension \cite{O44}\footnote{But our detailed examination of finite-size effects is, as far as we know, new.}. The transfer matrix from edge to edge of the strip, i.e., along the cylinder axis will be used \footnote{In this section $M$ denotes the length of the horizontal edge and $N$ the strip width, this notation is 
due to historical reasons. The final results are obviously covariant under the mutual exchange of $K_{1}$ with $K_{2}$.}. Thus we have an underlying translational, or cyclic, symmetry. The inter-row transfer matrix is the same as used in Sec.~\ref{subsec:cor}  - see Eq.~(\ref{eq:t1}). The intra-row matrix is  given by
\begin{equation}
\label{eq:dw1 }
V_{2} =  \exp\left( K_{2}\sum_{m=1}^{M}\sigma_{m}^{x}\sigma_{m+1}^{x}\right) \, ,
\end{equation}
where $\sigma_{M+1}^{x}=\sigma_{1}^{x}$, as required by the cyclic boundary conditions. 
The disordered state representing the free boundary, denoted $0$, may be taken as the state with all spins down in the $z$-direction. Then the partition function for the strip is
\begin{equation}
\label{ }
Z =  \langle 0 \vert \left(V_{2}V_{1}\right)^{N-1}V_{2}\vert0\rangle \, .
\end{equation}
The analogous quantity for the modified lattice is given by $Z^{\times}$ where
\begin{equation}
\label{ }
Z^{\times} =  \langle 0 \vert \left(V_{2}^{\times}V_{1}\right)^{N-1}V_{2}^{\times}\vert0\rangle \, ,
\end{equation}
the operator $V_{1}$ is defined by (\ref{eq:t1}) but notice that we omitted the factor in front of the exponential. The modified $V_{2}$ is given by
\begin{equation}
\label{ }
V_{2}^{\times} = \exp\left(K_{2}\sum_{m=1}^{M-1}\sigma_{m}^{x}\sigma_{m+1}^{x}-K_{2}\sigma_{M}^{x}\sigma_{1}^{x}\right) \, .
\end{equation}
The key to evaluating $Z$ and $Z^{\times}$ is to introduce a symmetrised transfer matrix in each case and then to use the Jordan-Wigner transformation to lattice Fermions. Define:
\begin{equation}
\label{ }
V^{\prime} = V_{1}^{1/2} V_{2} V_{1}^{1/2} \, ,
\end{equation}
and
\begin{equation}
\label{ }
\left(V^{\prime}\right)^{\times} = V_{1}^{1/2} V_{2}^{\times} V_{1}^{1/2} \, .
\end{equation}
Then, noting that $V_{1}\vert0\rangle = \textrm{e}^{MK_{1}^{*}} \vert0\rangle$, it follows that
\begin{equation}
\label{ }
\frac{Z^{\times}}{Z} = \frac{\bigl\langle0 \big\vert \bigl[ \left(V^{\prime}\right)^{\times} \bigr]^{N} \big\vert 0 \bigr\rangle}{\bigl\langle0\big\vert \bigl[ V^{\prime} \bigr]^{N} \big\vert 0\bigr\rangle} \, .
\end{equation}
The Jordan-Wigner transformation is given by (\ref{eq:t4}) and the corresponding commutation relations for the lattice Fermi operators are described in Sec. \ref{sec:theory}.A. In terms of lattice fermions we then to show that
\begin{eqnarray}
V_{1} & = & \exp\Biggl[ -K_{1}^{*}\left(\sum_{m=1}^{M}2f_{m}^{\dag}f_{m}-1\right) \Biggr] \, , \\
V_{2} & = & \exp\Biggl[ K_{2}\sum_{m=1}^{M-1}\left(f_{m}^{\dag}-f_{m}\right)\left(f_{m+1}^{\dag}+f_{m+1}\right) - K_{2} P_{M}\left(f_{M}^{\dag}-f_{M}\right)\left(f_{1}^{\dag}+f_{1}\right) \Biggr] \, ,
\end{eqnarray}
but for the term $P_{M}$ in $V_{2}$ above, both $V_{1}$ and $V_{2}$ are exponentials of quadratic forms in fermion operators. Moreover, $[V_{1},P_{M}]=0$ and $[V_{2},P_{M}]=0$. Thus, we can project onto the invariant subspaces of $P_{M}$ and consider:
\begin{equation}
\label{ }
V_{2}(\pm) = \exp\biggl[K_{2}\sum_{m=1}^{M-1} \left(f_{m}^{\dag}-f_{m}\right)\left(f_{m+1}^{\dag}+f_{m+1}\right) \mp K_{2} \left(f_{M}^{\dag}-f_{M}\right)\left(f_{1}^{\dag}+f_{1}\right) \biggr] \, ,
\end{equation}
and $V^{\prime}(\pm) = V_{1}^{1/2} V_{2}(\pm) V_{1}^{1/2}$. Then, using $P_{M}\vert0\rangle=\vert0\rangle$ the required ratio of partition functions becomes
\begin{equation}
\label{partitionfunctions}
\frac{Z^{\times}}{Z} = \frac{\langle0\vert \left( V^{\prime}(-) \right)^{N} \vert0\rangle}{\langle0\vert \left( V^{\prime}(+) \right)^{N} \vert0\rangle} \, .
\end{equation}
The evaluation of (\ref{partitionfunctions}) can be carried out with the technique of SML, where one uses lattice Fourier transformation
\begin{equation}
\label{ }
F(k) = M^{-1/2} \sum_{m=1}^{M} \textrm{e}^{-ikm}f_{m} \, ,
\end{equation}
with momenta $k$ restricted in two different sets depending on $\textrm{e}^{ikM}=\mp1$. Consideration of the translational symmetry of the original lattice Pauli spin operators makes the occurrence of these curious periodic and anti-periodic momenta reasonable. Then, by bringing in the pairing ideas of Nambu \cite{nambu} and of Anderson \cite{anderson}, the above quotient can be evaluated as a ratio of products
\begin{equation}
\label{Z_products}
\frac{Z^{\times}}{Z} = \prod_{j=1}^{M} \biggl[ \frac{g_{N}\left((2j-1)\pi/M\right)}{g_{N}\left(2(j-1)\pi/M\right)} \biggr]^{1/2} \, ,
\end{equation}
with the $2\pi$-periodic function $g_{N}(k)$ defined by
\begin{equation}
\label{g}
g_{N}(k) = \cosh N\gamma(k) + \sinh N\gamma(k) \cos\delta^{\prime}(k) \, .
\end{equation}
The detailed derivation of (\ref{Z_products}) from (\ref{partitionfunctions}) is not reported here but it can be carried out using the formalism developed in Ref. \cite{Abraham76}. The functions $\gamma$, $\delta^{\prime}$ and $\delta^{*}$ were introduced by Onsager as elements of a hyperbolic triangle in the Beltrami-Poincar$\acute{\textrm{e}}$ unit disk version of non-Euclidean geometry:
\begin{eqnarray}\nonumber
\cosh\gamma(k) & = & \cosh2K_{1}^{*}\cosh2K_{2}-\sinh2K_{1}^{*}\sinh2K_{2}\cos k , \\
\cosh2K_{1}^{*} & = & \cosh2K_{2}\cosh\gamma(k) - \sinh2K_{2}\sinh\gamma(k)\cos\delta^{*}(k) , \\
\cosh2K_{2} & = & \cosh2K_{1}^{*}\cosh\gamma(k) - \sinh2K_{1}^{*}\sinh\gamma(k)\cos\delta^{\prime}(k) , \nonumber
\end{eqnarray}
these are the hyperbolic cosine formulae for the Onsager hyperbolic triangle \cite{O44}, which should be supplemented by the hyperbolic sine formulae:
\begin{equation}
\label{ }
\frac{\sin\delta^{*}(k)}{\sin2K_{1}^{*}} = \frac{\sin\delta^{\prime}(k)}{\sin2K_{2}} = \frac{\sin k}{\sinh\gamma(k)} \, .
\end{equation}
These formulae are extremely useful for simplifying expressions, as should become apparent. The evaluation of the ratio of products may be made by first exponentiating (\ref{Z_products}): then we have
\begin{equation}
\frac{Z^{\times}}{Z} = \exp \Biggl[ \frac{1}{2} \sum_{j=1}^{M}\bigl( \ln g_{N}\left((2j-1)\pi/M\right) - \ln g_{N}\left(2(j-1)\pi/M\right) \bigr) \Biggr] \, .
\end{equation}
In order to use summation kernels to evaluate this difference as a contour integral, we need the analytic properties of $g_{N}(k)$ and, of course, its zeros and poles. The branch cuts from $\sinh\gamma(k)$ in $(\ref{g})$ does not contribute. Evidently, for $k\in\mathbb{R}$, $g_{N}(k)>0$ and, moreover, this property extends to an interval $|\textrm{Im}(k)|<\epsilon$, $\epsilon<\hat{\gamma}(0)$ where $\hat{\gamma}$ defined by analogy with $\gamma$ but with $K_{1}$ and $K_{2}$ interchanged. Then we can write
\begin{equation}
\label{ }
\frac{Z^{\times}}{Z} = \exp\biggl[ \frac{1}{2} \frac{1}{2\pi i} \oint_{\mathfrak{C}}\textrm{d}k \, i M \left( \frac{1}{\textrm{e}^{ikM}-1} + \frac{1}{\textrm{e}^{ikM}+1} \right) \ln g_{N}(k) \biggr] \, ,
\end{equation}
where the contour $\mathfrak{C}$ is shown in Fig.\ref{fig_contour}.
Simplifying this gives
\begin{equation}
\label{ }
\frac{Z^{\times}}{Z} = \exp\biggl[ \frac{M}{4\pi i} \oint_{\mathfrak{C}}\textrm{d}k \, \frac{1}{\sin Mk} \ln g_{N}(k) \biggr] \, .
\end{equation}
Using the even character of $\ln g_{N}(k)$, this may be written as

\begin{equation}
\label{ }
\frac{Z^{\times}}{Z} = \exp\biggl[ \frac{M}{\pi} \int_{-\pi+i\epsilon}^{\pi+i\epsilon}\textrm{d}k \, \frac{\textrm{e}^{iMk}}{1-\textrm{e}^{2iMk}} \ln g_{N}(k) \biggr] \, ,
\end{equation}
we now expand the integrand using the geometric series, reorder summation and integration by standard theorems to get
\begin{equation}
\label{ }
\frac{Z^{\times}}{Z} = \exp\biggl[ \sum_{j=0}^{\infty} \frac{M}{\pi} \int_{-\pi+i\epsilon}^{\pi+i\epsilon}\textrm{d}k \, \textrm{e}^{i(2j+1)Mk} \ln g_{N}(k) \biggr] \, ,
\end{equation}
now integrate by parts
\begin{equation}
\label{ }
\frac{Z^{\times}}{Z} = \exp\biggl[ -\sum_{j=0}^{\infty} \frac{1}{2j+1} \frac{1}{\pi i} \int_{-\pi+i\epsilon}^{\pi+i\epsilon}\textrm{d}k \, \textrm{e}^{i(2j+1)Mk} \, \frac{g_{N}^{\prime}(k)}{g_{N}(k)} \biggr] \, .
\end{equation}
The remaining part of the evaluation is to find the zeros of $g_{N}(k)$. Introducing the conformal transformation $k=i\hat{\gamma}(u)$ rather conveniently does this, since
\begin{equation}
\label{ }
g_{N}\left(i\hat{\gamma}(u)\right) = \cos Nu + \frac{\sin Nu}{\sin u} \frac{\cosh2K_{1}^{*}\cos u - \cosh2K_{2}}{\sinh2K_{1}^{*}} \, .
\end{equation}
The problem of evaluating zeros of $g_{M}\left(\hat{\gamma}(u)\right)$ can then be reduced to finding the solutions of
\begin{equation}
\label{ }
\textrm{e}^{2iNu} = \textrm{e}^{2i\widehat{\delta^{*}}(u)} \, ,
\end{equation}
where the angle $\widehat{\delta^{*}}$ is derived from $\delta^{*}$ by interchanging $K_{1}$ and $K_{2}$. Of particular interest is the sub-critical region. If $N<\kappa$, with $\kappa \equiv \textrm{d}\widehat{\delta^{*}}(\omega)/\textrm{d}\omega \vert_{\omega=0}$, there are $N$ real solutions $u_{j}$, $j=1,\dots,N$ such that for each such $j$ there is a solution in the open interval $\left( \pi(j-1)/N, \pi j/N \right)$. We now select the zeros of $g_{N}(k)$ in the upper half plane, and get
\begin{equation}
\label{ }
\frac{Z^{\times}}{Z} = \exp\biggl[ -\sum_{\ell=1}^{N} \sum_{j=0}^{\infty} \frac{2}{2j+1} \textrm{e}^{-(2j+1)M\hat{\gamma}(u_{\ell})} \biggr] \, ,
\end{equation}
the sum can be identified and carried out explicitly, giving
\begin{equation}
\label{Ztanhhatversion}
\frac{Z^{\times}}{Z} = \prod_{\ell=1}^{N}\tanh\left(\frac{M\hat{\gamma}(u_{\ell})}{2}\right) \, .
\end{equation}
On the other hand, if $N>\kappa$ and, of course $K_{1}^{*}<K_{2}$ there is a single imaginary solution for $u$ in the upper half plane. What is happening is that two solutions $\pm u_{1}$ for $N>\kappa$ (if $u$ is a solution, then so is $-u$) coalesce at the origin when $N=\kappa$ and then go onto the imaginary axis as $\pm i v_{1}$, for $N>\kappa$, a bifurcation phenomenon, with
\begin{equation}
\label{gamma_iv}
\hat{\gamma}(iv_{1}) = 2\sinh2K_{2}^{*} \sinh\gamma(0) \, \textrm{e}^{-N\gamma(0)} + \mathcal{O}\left( \textrm{e}^{-2N\gamma(0)} \right) \, .
\end{equation}
We then find the asymptotic form
\begin{equation}
\label{asymptotics}
\frac{Z^{\times}}{Z} = M\sinh2K_{2}^{*}\sinh\gamma(0) \, \textrm{e}^{-N\gamma(0)} + \mathcal{O}\left( \textrm{e}^{-2N\gamma(0)} \right) \, .
\end{equation}
This is very satisfactory, since with the cyclic boundary conditions as indicated there are $M$ precisely equivalent translates of any configuration, hence the factor $M$ which gives a Boltzmann entropy $\ln M$ in units of $k_{B}T$. The correct Boltzmann weight for a domain wall is thus
\begin{equation}
\label{exactweight}
w = \sinh2K_{2}^{*}\sinh\gamma(0)\textrm{e}^{-N\gamma(0)} \, .
\end{equation}
There is a surface tension of $\gamma(0)$, in agreement with Onsager \cite{O44}, and a point tension\footnote{The excess free energy (in $k_{B}T$ units) takes the form $\mathcal{F}=-\ln M + N\gamma(0) - \ln\bigl[ \sinh2K_{2}^{*}\sinh\tau\bigr]$. The first term $-\ln M$ is an entropic contribution due to the fact that we can locate the domain wall in $M$ positions which are equivalent  under translation invariance, $N\hat{\gamma}(0)$ is the energy cost of an unpinned domain wall, $\tau_{p}= -2^{-1}\ln\bigl[ \sinh2K_{2}^{*}\sinh\tau\bigr]$ is the point tension originated at each anchoring point.} (the $d=2$ analogue of line tension in $d=3$) having the value $\tau_{p}$, where
\begin{equation}
\label{pointtension}
\tau_{p} = -2^{-1}\ln\bigl[ \sinh2K_{2}^{*}\sinh\tau\bigr] \, .
\end{equation}
This is, we believe, a new result. Note that it has a logarithmic singularity at the critical point. The reader who is acquainted with the classical results of SML would more than likely choose the above method to evaluate the domain wall weight $w$.

Another derivation follows, one which makes the origin of the product of hyperbolic tangents, the function $\gamma(u)$ and the particular choice of the $u_{j}$ clearer, that is, one which makes the \emph{physical} origin of the structure more obvious. If we consider transfer along the strip as in Fig.\ref{fig:1}, that is, exactly what took us to the formula for the pair correlation function for spins in the edge of a strip, we have:
\begin{equation}
\label{ }
\frac{Z^{\times}}{Z} = \frac{\textrm{Tr}\Bigl[ \left(V^{\prime}\right)^{N}P_{M} \Bigr]}{\textrm{Tr}\Bigl[ \left(V^{\prime}\right)^{N} \Bigr]} \, ,
\end{equation}
and
\begin{equation}
\label{Z_tanh}
\frac{Z^{\times}}{Z} = \prod_{k}\tanh\left( \frac{N\gamma(k)}{2} \right) \, .
\end{equation}
Here, we have
\begin{eqnarray}
V_{1} & = & \exp\biggl[ -K_{1}^{*}\sum_{m=1}^{M}\left( 2f_{m}^{\dag}f_{m}-1 \right) \biggr] \, , \\
V_{2} & = & \exp\biggl[ K_{2}\sum_{m=1}^{M-1}\left( f_{m}^{\dag}-f_{m} \right)\left( f_{m+1}^{\dag}+f_{m+1} \right) \biggr] \, .
\end{eqnarray}
Notice the range of summation in $V_{2}$, as required by the free-edged strip. Now it turns out that symmetrised form $V$ can be diagonalised \cite{DBA:71}, in the form
\begin{equation}
\label{ }
V = \exp\biggl[ -\sum_{k}\gamma(k)\left(X^{\dag}(k)X(k)-1/2\right)\biggr] \, .
\end{equation}
The $\bigl\{ X(k), X^{\dag}(k)\bigr\}$ are Fermi operators with vacuum $\vert\Phi\rangle$ which therefore satisfies $X(k) \vert\Phi\rangle=0$; moreover, $P_{M}\vert\Phi\rangle=\vert\Phi\rangle$. Since we have
\begin{equation}
\label{ }
P_{M} X^{\dag}(k_{1}) \dots X^{\dag}(k_{n}) \vert\Phi\rangle = (-1)^{n} X^{\dag}(k_{1}) \dots X^{\dag}(k_{n}) \vert\Phi\rangle \, ,
\end{equation}
and $n_{k}=X^{\dag}(k)X(k)$ is the density operator, it follows that
\begin{equation}
\label{ }
\frac{Z^{\times}}{Z} = \prod_{k}\frac{\sum_{n_{k}=0}^{1}\Bigl[ \textrm{e}^{-\left(N\gamma(k)+i\pi\right) n_{k}} \Bigr]}{\sum_{n_{k}=0}^{1}\Bigl[ \textrm{e}^{-N\gamma(k) n_{k}} \Bigr]} \, ,
\end{equation}
and thus (\ref{Z_tanh}) follows, where the $k$ are given by
\begin{equation}
\label{ }
\textrm{e}^{2iMk} = \textrm{e}^{2i\delta^{*}(k)} \, .
\end{equation}
This is exactly as we have derived in (\ref{asymptotics}), (\ref{exactweight}) and (\ref{pointtension}), provided one remembers to interchange the $K_{j}$, $j=1,2$, and also $N$ and $M$ (compare (\ref{Z_tanh}) and (\ref{Ztanhhatversion})).

%====================================
%================= SECTION 2.D
%====================================
\subsection{Asymptotic degeneracy}
\label{subsec:asymptotic_degeneracy}
The transfer matrix acting parallel to the strip axis (see Fig.\ref{fig:1}) has a unique maximal eigenvector $\vert\Phi\rangle$ with eigenvalue $\Lambda_{0}$, which is also the vacuum for Fermi creation operators $X(k)$: $X(k)\vert\Phi\rangle=0$. For $T<T_{c}$ and a strip width $M$ satisfying $M>\textrm{d}\delta^{*}(\omega)/\textrm{d}\omega \vert_{\omega=0}$, we have a mode with a purely imaginary wavenumber, excited by the creation operator $X^{\dag}(iv)$. Its eigenvalue is $\Lambda_{0} \, \textrm{e}^{\gamma(iv)}$. So from (\ref{gamma_iv}), it is asymptotically degenerate with $\vert\Phi\rangle$. Now $\vert\Phi\rangle$ is strictly non-degenerate for $M<\infty$. Since $[V,P_{M}]=0$, $\vert\Phi\rangle$ must be simultaneous eigenvector of $P_{M}$, so since $P_{M}^{2}=1$,
\begin{equation}
\label{ }
P_{M} \vert\Phi\rangle = \pm \vert\Phi\rangle \, .
\end{equation}
In fact, $P_{M} \vert\Phi\rangle=\vert\Phi\rangle$. Because $P_{M}\sigma_{1}^{x}P_{M}=-\sigma_{1}^{x}$ we find
\begin{equation}
\label{ }
\langle\Phi \vert \sigma_{1}^{x} \vert\Phi\rangle = 0 \, .
\end{equation}
Thus there is \emph{never} long range order in a strip of finite width. Equally well, we have $\langle\Phi\vert X(iv) \sigma_{1}^{x} X^{\dag}(iv) \vert\Phi\rangle=0$. On examining (\ref{eq:t22}) in the limit $M\rightarrow\infty$, we see that the first term no longer decays to zero as $n\rightarrow\infty$; in fact, it decays to $m_{e}^{2}$. This is because the emergent length scale in (\ref{eq:t22}), namely $\xi = \textrm{e}^{M\hat{\gamma}(0)}/(2\sinh\hat{\gamma}(0)\sinh2K_{1}^{*})$, diverges as $M\rightarrow\infty$. The second term becomes an integral and displays a correlation length $1/\gamma(0)$; thus it vanishes in the ``limit'' of meso-scale modeling. It is natural to specify putative ordered states (which are \emph{not} eigenstates of $V$)
\begin{equation}
\label{ }
\vert \pm \rangle = 2^{-1/2} \left( 1\pm X^{\dag}(iv) \right) \vert\Phi\rangle \, ,
\end{equation}
which evidently have the property $P_{M}\vert\pm\rangle=\vert\mp\rangle$. The edge magnetization for $\vert+\rangle$ is
\begin{equation}
\label{ }
\langle+\vert \sigma_{1}^{x} \vert+\rangle = \textrm{Re} \langle\Phi\vert \sigma_{1}^{x} X^{\dag}(iv) \vert\Phi\rangle \, .
\end{equation}
In this connection, there is an analogous formulation for the spontaneous magnetization of the bulk. The evaluation of the associated matrix element in that case is a true \emph{tour de force}, carried out by Yang \cite{Yang}. The eigenvectors for $k=iv$ are obtained by noting this substitution in (\ref{eq:t16}) and (\ref{eq:t17}). The resulting mode is indeed a surface state in the Fermi lattice language.

%====================================
%================= SECTION 3
%====================================
\section{Numerical simulations}
\label{sec:num}
In order to test our predictions based on the extended Fisher-Privman theory \cite{privman_fisher} we have performed a series of Monte Carlo simulations of the Ising model.

%====================================
%================= SECTION 3.A
%====================================
\subsection{Numerical method and observables}
\label{subsec:1}
We consider the Ising model on  a  square lattice in two dimensions (2D) and on a  simple cubic lattice in three dimensions (3D) with the lattice spacing $\ell=1$ defined via the Hamiltonian
\be
\label{eq:Ham}
H = - J \sum_{\la i,j \ra}  s_{i}  s_{j} \, ,
\ee
where $s_{i}=\pm 1$ denotes the spin variable. The parameter $J$, which we set equal to $1$, is the spin-spin coupling constant and the sum $\la i,j \ra$ is taken over all nearest neighbor pairs of sites $i$ and $j$ on the lattice. The total number of spins of the lattice is given by $N_{s}$. We shall consider different geometries, which will be specified later. For all geometries we assume open boundary conditions (OBC) in which the spins are free at the boundaries. For the square lattice, the critical value of the coupling constant $K=\beta J$, where $\beta=1/(k_{B}T)$, is given by $K_{c} = (1/2)\ln(1+\sqrt{2}) \approx 0.440687$ \cite{O44}. Various estimations are available for $D=3$ \cite{RCW}; $K_{c}(D=3) \approx 0.2216544(3) \approx K_{c}(D=2)/2$.

We perform numerical simulation using a hybrid algorithm. One Monte Carlo step consists of one update of Wolf cluster and $N_{s}/4$ Metropolis updates of randomly selected spins. We use standard definitions~\cite{LB,BinderLujiten} for the thermodynamic quantities: the magnetization per spin is
\be
m=\frac{1}{N_{s}} \biggl\langle \Big\vert \sum \limits_{\{j \}}\sigma_{j} \Big\vert \biggr\rangle = \frac{1}{N_{s}} \la M\ra \, ,
\ee
where the sum $\{ j\}$ is taken over all spins of the system, the energy per spin is given by
\be
e=-\frac{1}{N_{s}} \biggl\langle \sum \limits_{\{i,j \}}\sigma_{i}\sigma_{j}  \biggl\rangle = \frac{1}{N_{s}} \la E\ra \, ,
\ee
the heat capacity is
\be
C=\beta^{2}\left(\la E^{2} \ra-\la E \ra^{2} \right)/N_{s} \, ,
\ee
and the magnetic susceptibility is
\be
\label{eq:chi_def}
\chi=\beta^{2}\left(\la M^{2} \ra-\la M \ra^{2} \right)/N_{s} \, ,
\ee
where $\la E^{2} \ra= \bigl\langle ( \sum_{\{i,j \}}\sigma_{i}\sigma_{j} )^{2} \bigl\rangle$ and $\la M^{2} \ra= \bigl\langle ( \sum_{\{j \}}\sigma_{j} )^{2} \bigl\rangle$. In the above definitions   $\la \dots \ra$ denotes the thermodynamic average over system states.

%====================================
%================= SECTION 3.B
%====================================
\subsection{1D array of squares.}
\label{subsec:2}
\begin{figure}[h]
\includegraphics[width=0.90\textwidth]{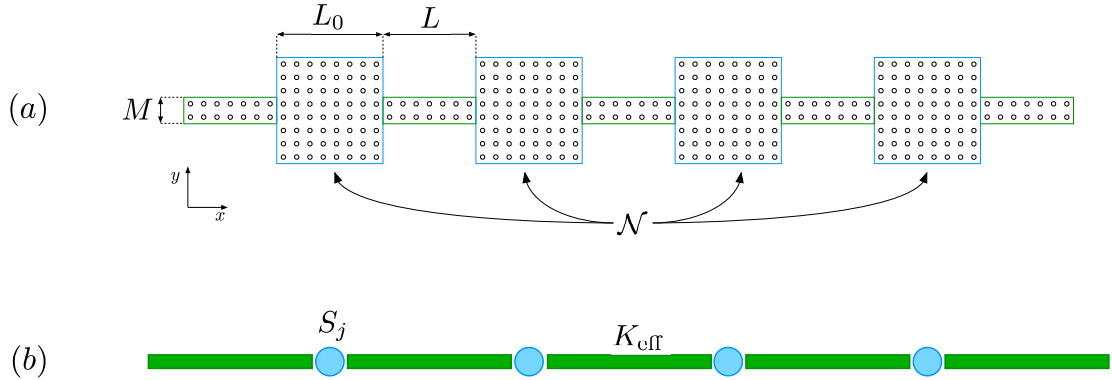}
\caption{($a$) Geometry of a 1D array of $\mathcal{N}$ squares of size $L_{0}$ connected by strips (channels) of length $L$ and thickness $M$. ($b$) The equivalent 1D system, that consists of $\mathcal{N}$ coarse grained spin variables $S_{j}$ connected by bonds with effective interaction $K_{\textrm{eff}}$.}
\label{fig:geom}
\end{figure}

Our lattice  network model of hyper-cubic Ising boxes connected by Ising strips (see Sec.~\ref{subsec:dw}) does not exhibit a phase transition in one dimension.  For the 1D network, however, we can use the exact form for the correlation function $G(x)=\langle S_{i} S_{j} \rangle$ between the boxes $S_{i}$ and $S_{j}$, separated by $x/L$ sites of the effective lattice with $x=i-j$; the latter reads
\begin{equation}
\label{Correlation_linear}
G(x) \simeq (m_0^*)^2(\tanh K_{\textrm{eff}})^{| x/L|} \, ,
\end{equation}
where $K_{\textrm{eff}}$ is the effective coupling interaction, which from (\ref{Boltzmann}) admits the neat expression $\tanh K_{\textrm{eff}}=t^{L}$. In order to test this prediction, we take the system that  consists of $\mathcal{N}$ squares of the size $L_{0}\times L_{0}$ (each  square contains $L_{0}^{2}$ spins) connected by strips of the length $L$ and width $M$ (the number of spins in the strip is equal to $L \times M$) see Fig.~\ref{fig:geom}(a). The system is periodic in $x$  direction,  the 1st and the $\mathcal{N}$th squares are connected forming a ring. The total number of spins in the system is $N_{s}=\mathcal{N}(L_{0}^{2}+LM)$. We have performed Monte Carlo simulations for the system of $\mathcal{N}=100$ squares of size $L_{0}=100$ connected by channels of length $L=100$ and various widths $M=4,6,10,20,30,40$.

\begin{figure}[h]
\includegraphics[width=0.8\textwidth]{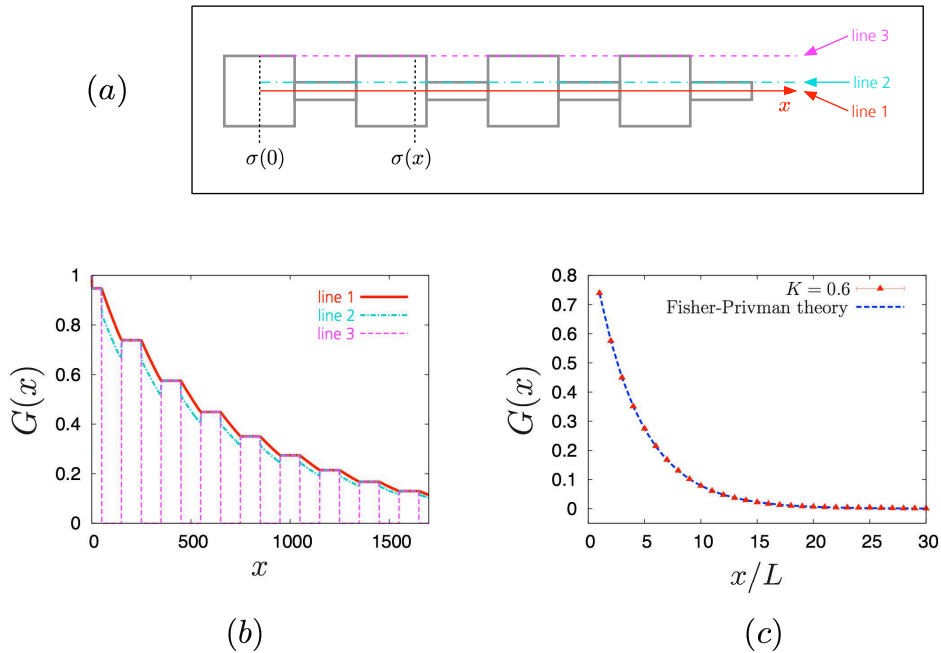}
%\vspace*{0.5cm}
%\begin{minipage}{0.45\textwidth}
%\includegraphics[width=\textwidth]{figure9b.eps}
%\end{minipage}
%\begin{minipage}{0.5\textwidth}
%\includegraphics[width=0.9\textwidth]{figure9c.eps} 
%\end{minipage}
\caption{($a$) The scheme for the computation of the spin-spin correlation function $G(x)$ along  three different lines: line 1 at the middle of the channel (red solid line), line 2 at the side of the channel (cyan dash-dotted line) and line 3 at the side of the square (magenta dashed line); ($b$) Spin-spin correlation function $G(x)$ for the 1D array of $\mathcal{N}=100$ squares of a side length $L_{0}=100$ as a function of the distance $x$ along the three lines: the line passing through center of the channel (red solid line), through the side of the channel (cyan dash-dotted line) and through the side of the square (magenta dashed line) for $K=0.6$; the channel length is $L=100$ and the width is $M=10$. ($c$) The plateau values of $G(x)$ computed along the middle line of the channel as a function of $x/L$ (symbols) follow  the 1D Ising correlation function law given by Eq.~(\ref{Correlation_linear}) (dashed blue line) with $m_0^*=0.97$.}
\label{fig:1d_100-100_corr1}
\end{figure}

The spin-spin correlation function $G(x)= \la \sigma(0) \sigma(x)\ra$ is computed in the  $x$ direction along three different lines, as shown in Fig.~\ref{fig:1d_100-100_corr1}(a). The  horizontal coordinate $x=0$  of the first spin $\sigma(0)$ is  always at the center of the square box. For the vertical coordinate $y$ of both spins we consider three cases: the centers of the squares, the sides of the channels, and the sides of the squares, denoted respectively with the lines $1$, $2$ and $3$ of Fig.\ref{fig:1d_100-100_corr1}(a). Let us note that, the correlations along the edges of the squares exist only if the second spin $\sigma(x)$ is located within a square.

In Fig.~\ref{fig:1d_100-100_corr1}(b) we plot the spin-spin correlation function along these lines for a channel of width $M=10$ and for the coupling $K=0.6$. One can see that $G(x)$ stays constant within the squares and depends only on the mutual distance between the latter, which supports the crucial assumption for derivations  in Sec.~\ref{subsec:dw} that the boxes  are  ordered. Fig.~\ref{fig:1d_100-100_corr1}(c) shows  the values of   plateaux from Fig.~\ref{fig:1d_100-100_corr1}(b) plotted as function of $x/L$ together with the theoretical prediction given by Eq.~(\ref{Correlation_linear}). Perfect agreement is obtained for  $m_0^*=0.97$ corresponding to the spontaneous magnetisation at $K=0.6$.
 
\begin{figure}[h]
\includegraphics[width=0.6\textwidth]{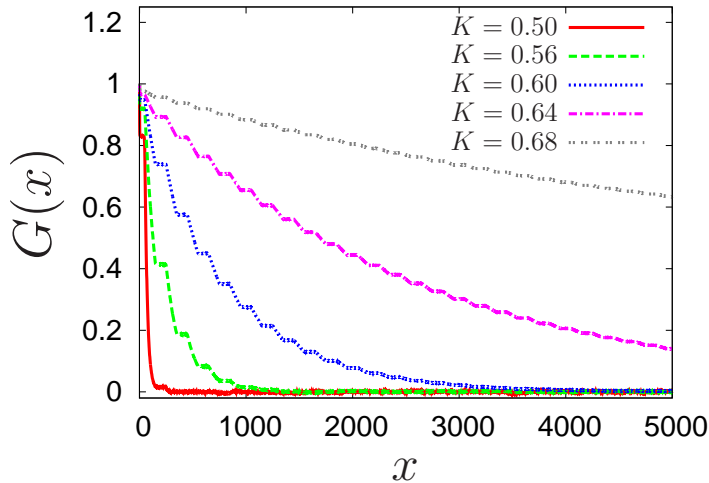}
\caption{
Spin-spin correlation function $G(x)$ along the centers of the squares (line 1) as a function of the distance $x$ for the same system as in Fig.~\ref{fig:1d_100-100_corr1} and various couplings, $K=0.5,0.56,0.6,0.64,0.68$.}
\label{fig:1d_100-100_corr2}
\end{figure}
\begin{figure}[h]
\mbox{
\includegraphics[width=0.49\textwidth]{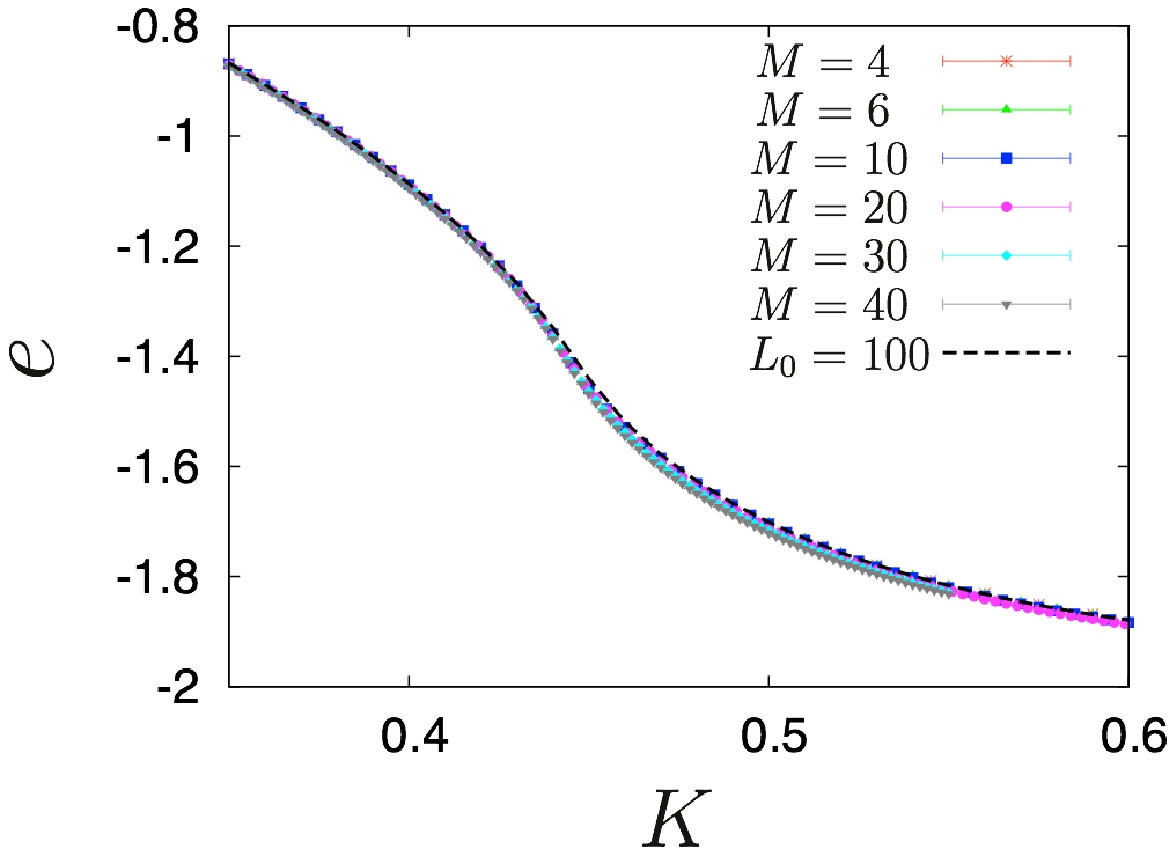}
\includegraphics[width=0.49\textwidth]{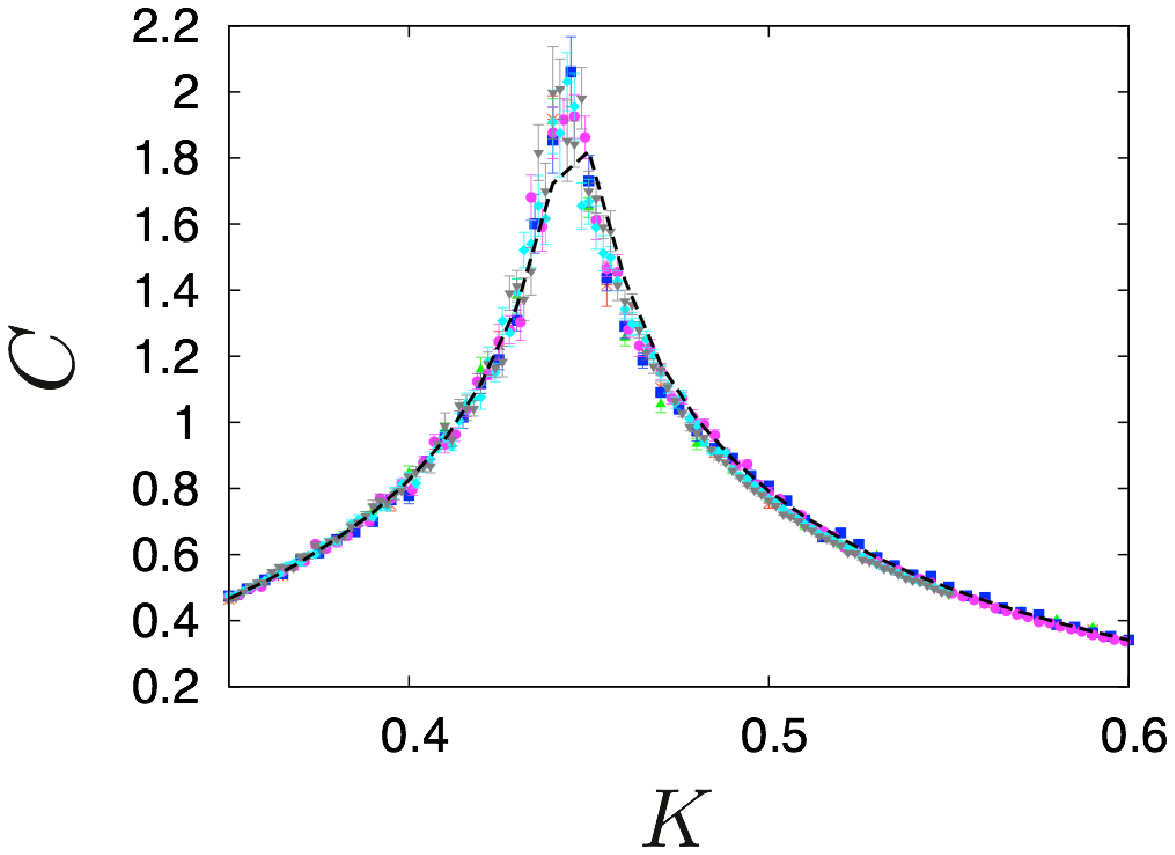}}
\mbox{
\includegraphics[width=0.49\textwidth]{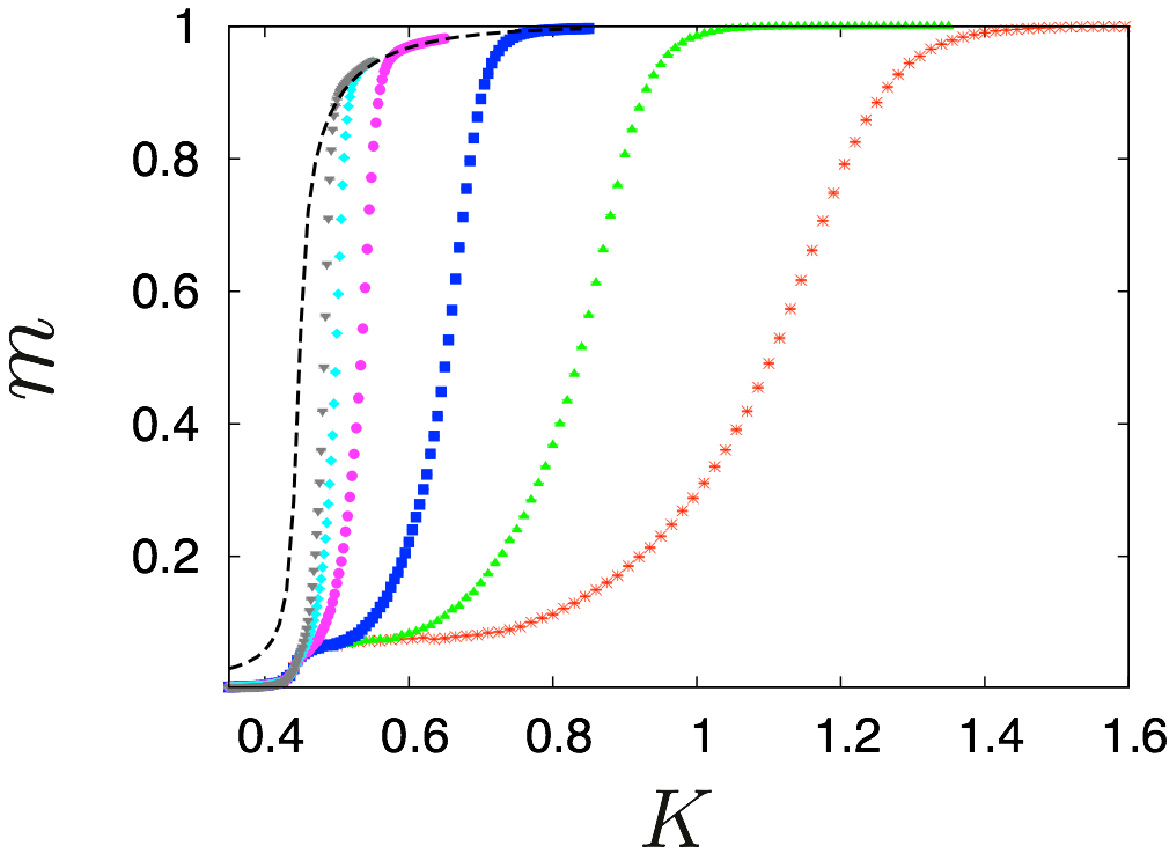}
\includegraphics[width=0.49\textwidth]{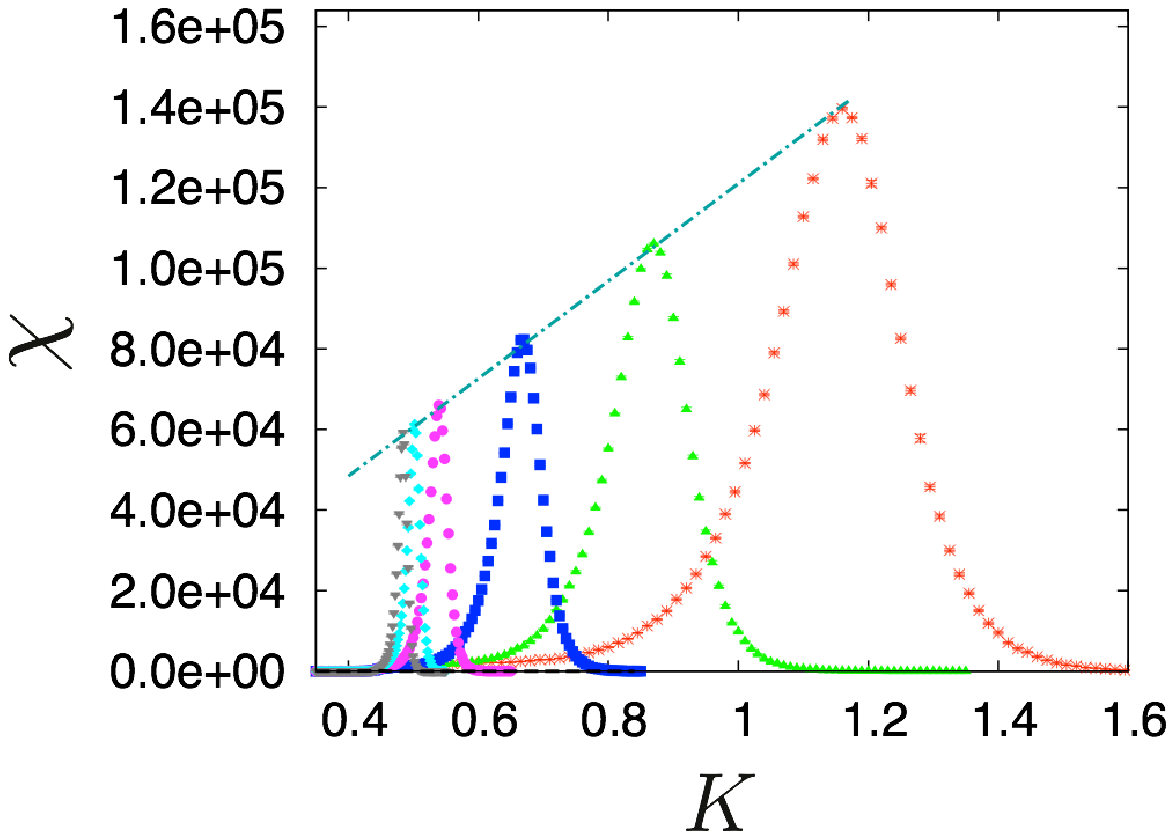}}
\caption{Thermodynamic quantities for a 1D array of $\mathcal{N}=100$ squares of size $L_{0}=100$ connected (in periodic way) by strips of size $100 \times M$ as a function of the coupling $K$: energy per spin $e$ (upper panel, left), heat capacity $C$ (upper panel, right), magnetization per spin $m$ (lower panel, left) and magnetic susceptibility $\chi$ (lower panel, right). Results for a single OBC  square with  $L_{0}=100$ are plotted by black dashed line for comparison (except for the plot of $\chi$).}
\label{fig:1d_100-100}
\end{figure}

In Fig.~\ref{fig:1d_100-100_corr2} we plot $G(x)$ for the same system as in Fig.~\ref{fig:1d_100-100_corr1} for several values of the coupling constant $K$. We can see that already for $K=0.5$ the spins within the first square are correlated, note that $G(x)>0.8$ for $x<50$. The spatial extent of the correlations grows by increasing the coupling $K$, and ultimately, the correlations spread across the whole system by further increasing of $K$. This feature can be linked to the behaviour of thermodynamic quantities as functions of $K$, which is presented  in Fig.~\ref{fig:1d_100-100}.

For comparison, we plot in this figure also the results for the single OBC square of the size $L_{0}=100$  (black dashed line). One can see that for the studied system sizes the energy-related quantities such as the energy per spin $e$ and the heat capacity for the single square and for the 1D array of coupled squares are almost identical; see Fig.~\ref{fig:1d_100-100}(a) and Fig.~\ref{fig:1d_100-100}(b). The peak in the heat capacity of a single PBC square indicates the rounded  2D continuous order-disorder phase transition. In contrast, the  magnetization-related quantities such as the magnetization per spin $m$ and the magnetic susceptibility exhibit a rounded transition at a value of $K_{c}(M) > 0.45$, which for this coupled system depends on the width of connecting channel $M$; see Fig.~\ref{fig:1d_100-100}(c) and Fig.~\ref{fig:1d_100-100}(d). The dash-dotted line in Fig.~\ref{fig:1d_100-100}(d) shows that the location of this rounded transition, as indicated by the maximum of the susceptibility $\chi_{\max}$, grows linearly with $K$ as the width of the channel $M$ is decreased. For the system of Fig.~\ref{fig:1d_100-100_corr2}, i.e., for $M=10$,  the maximum of the magnetic susceptibility $\chi$ occurs at  $K\approx 0.66$. Around this maximum the spins in the whole 1D array  become correlated as can be inferred  
from the behaviour of $G(x)$ shown in this figure.  This is a manifestation of  finite-size effects on the order-disorder transition in 2D systems~\cite{FSE}. The (pseudo) critical coupling is shifted to the higher values of $K$ with respect to $K_c(D=2)\approx 0.440687$ of the bulk  2D Ising model, this shift depends both on on the geometry and the size of the system.

We have checked how the size of constituents of the array influences both the heat capacity and the magnetic susceptibility. We have found that enlarging the connecting volumes leads to the increase of the maximum of the heat capacity, while the location of the peak remains practically unchanged. We observe that making the connecting channel shorter does not influence the heat capacity $C$. On contrary, the peak of the susceptibility become only slightly larger for larger boxes but  shortening the channel length leads to the shift of the peak of $\chi$ to regions of smaller $K$.

From Fig.~\ref{fig:1d_100-100}  we can conclude that the rounded ordering transition in the array of coupled volumes occurs in two stages. At the first stage spins in every square become ordered, this process takes place about the (pseudo) critical point $K_{c,L_0} \simeq 0.45$ for an isolated OBC square of size $L_{0}$ and does not depend on the geometry of the connecting channels. This first stage is responsible for the peak of the heat capacity. This rounded phase transition corresponds to a system with the same spatial dimension (and universality class) as  the coupled volumes, for the case at hand the volumes are actually two-dimensional entities. However right at this transition point different squares stay uncorrelated. As we increase the coupling parameter $K$, at a certain value $K_{c}^{*}(M,L)$ that depends on the size $L,M$ of channels, different squares become ordered. In correspondence of this second (rounded) transition the magnetization tends to unity and the magnetic susceptibility of the system reaches its maximum value.

%====================================
%================= SECTION 3.C
%====================================
\subsection{2D arrays of squares.} 
\begin{figure}[h]
\includegraphics[width=0.90\textwidth]{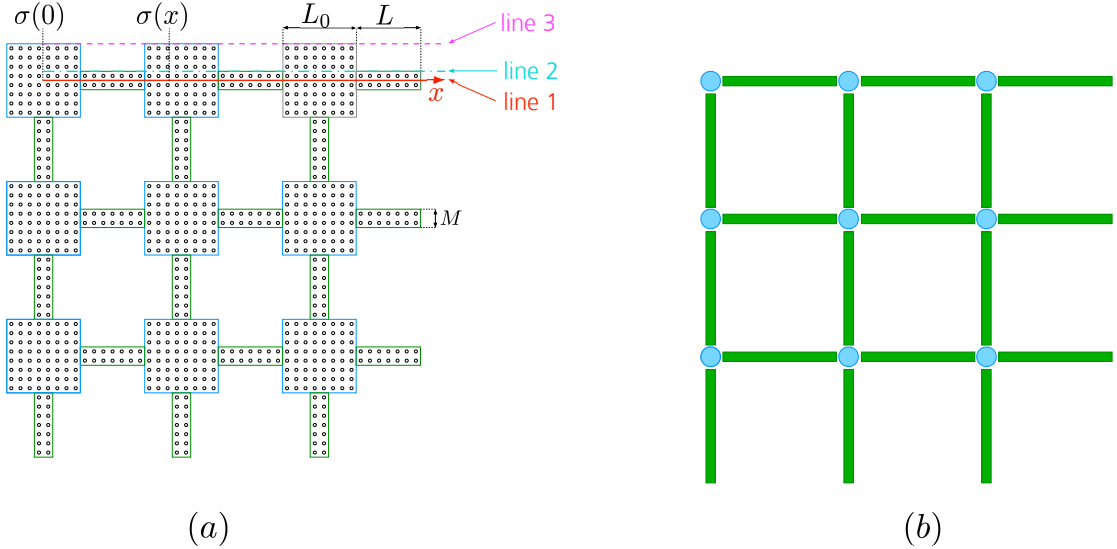}
\caption{($a$) Geometry of a 2D array of a linear size $\mathcal{N}$ consisting of $\mathcal{N}^{2}$ squares of size $L_{0}$ connected by strips (channels) of the length $L$ and the width $M$. The scheme for computation of the spin-spin correlation function $G(x)=\la\sigma(0)\sigma(x)\ra$ for three different lines: line 1 at the middle of the channel, (red solid line), line 2 at the side of the channel (cyan dash-dotted line), and line 3 at the side of the square (magenta dashed line). ($b$) Geometry of the equivalent network model.}
\label{fig:geom2d}
\end{figure}

\begin{figure}[h]
\mbox{
\includegraphics[width=0.49\textwidth]{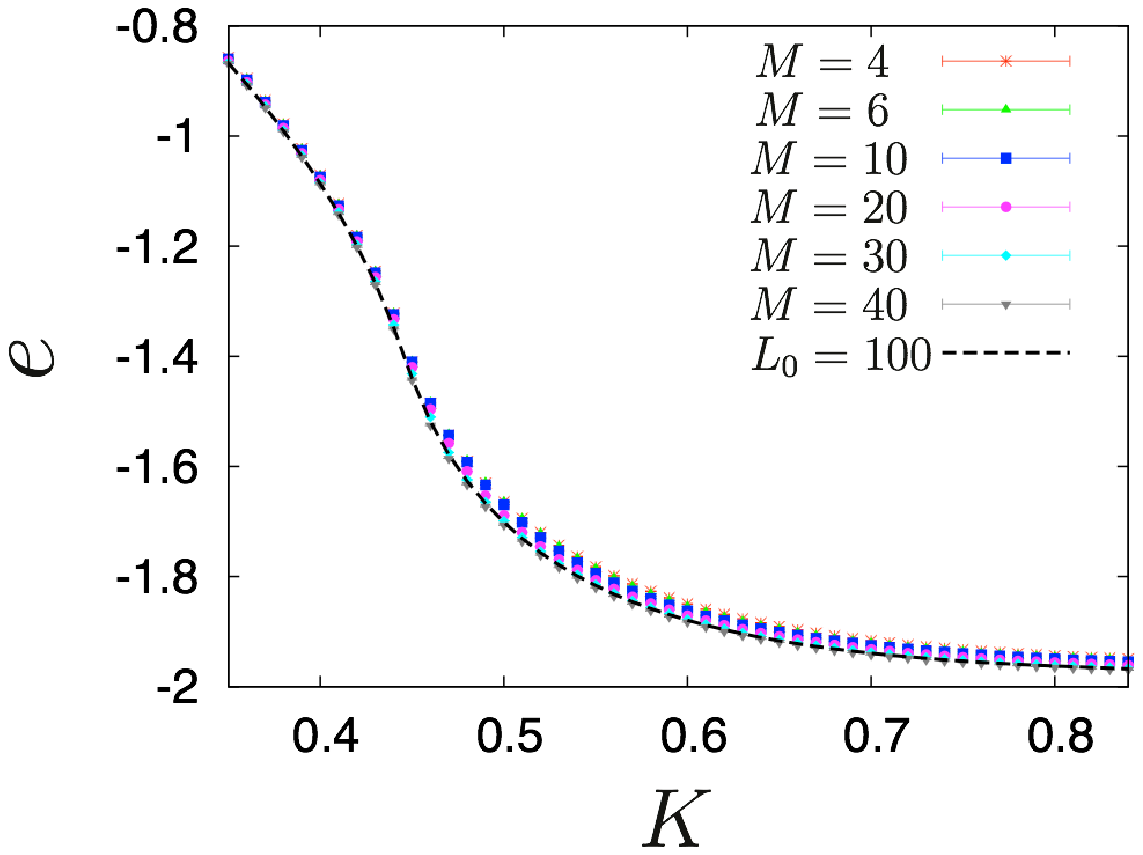}
\includegraphics[width=0.49\textwidth]{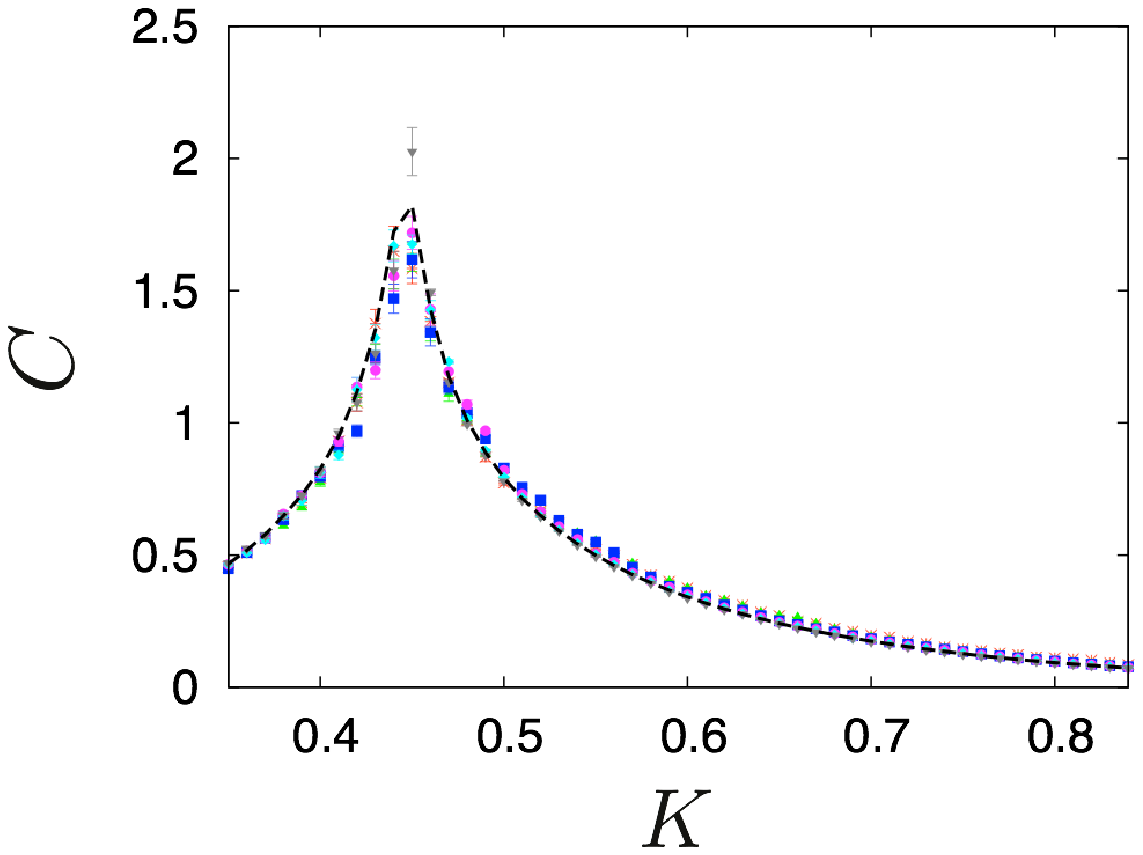}}
\mbox{
\includegraphics[width=0.49\textwidth]{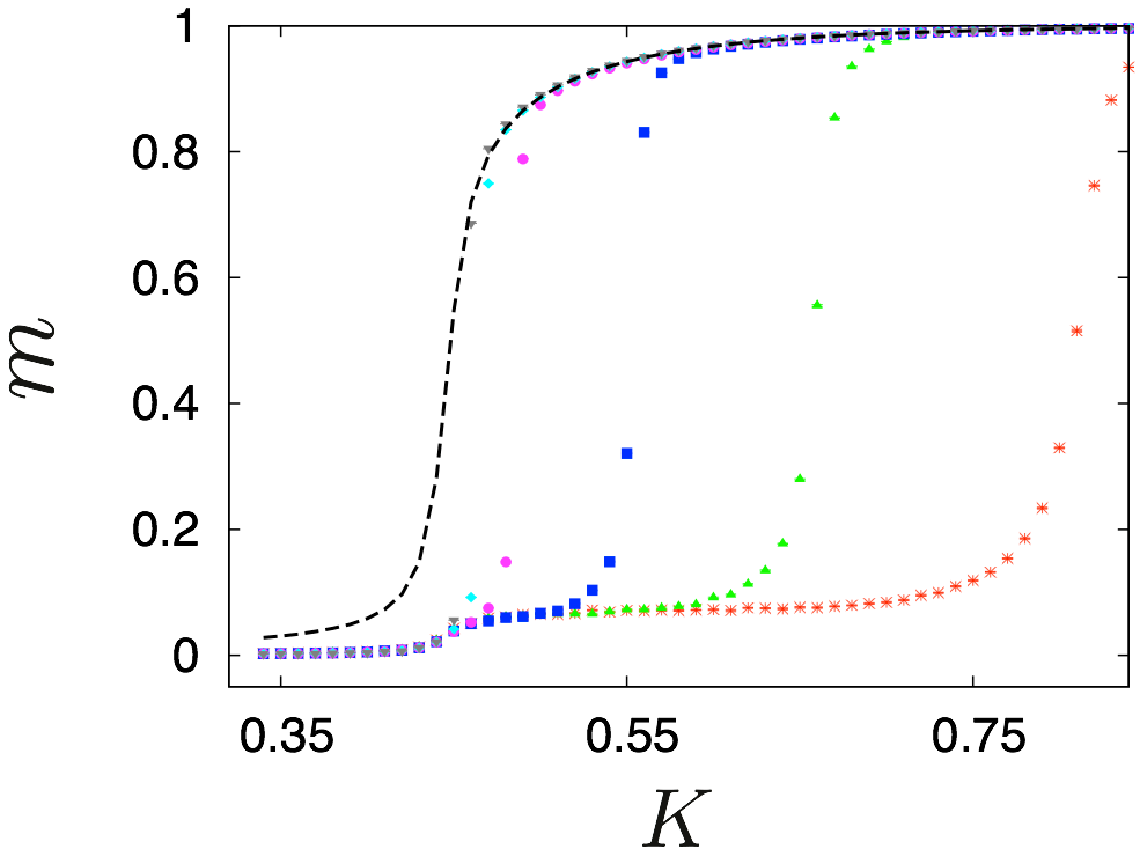}
\includegraphics[width=0.49\textwidth]{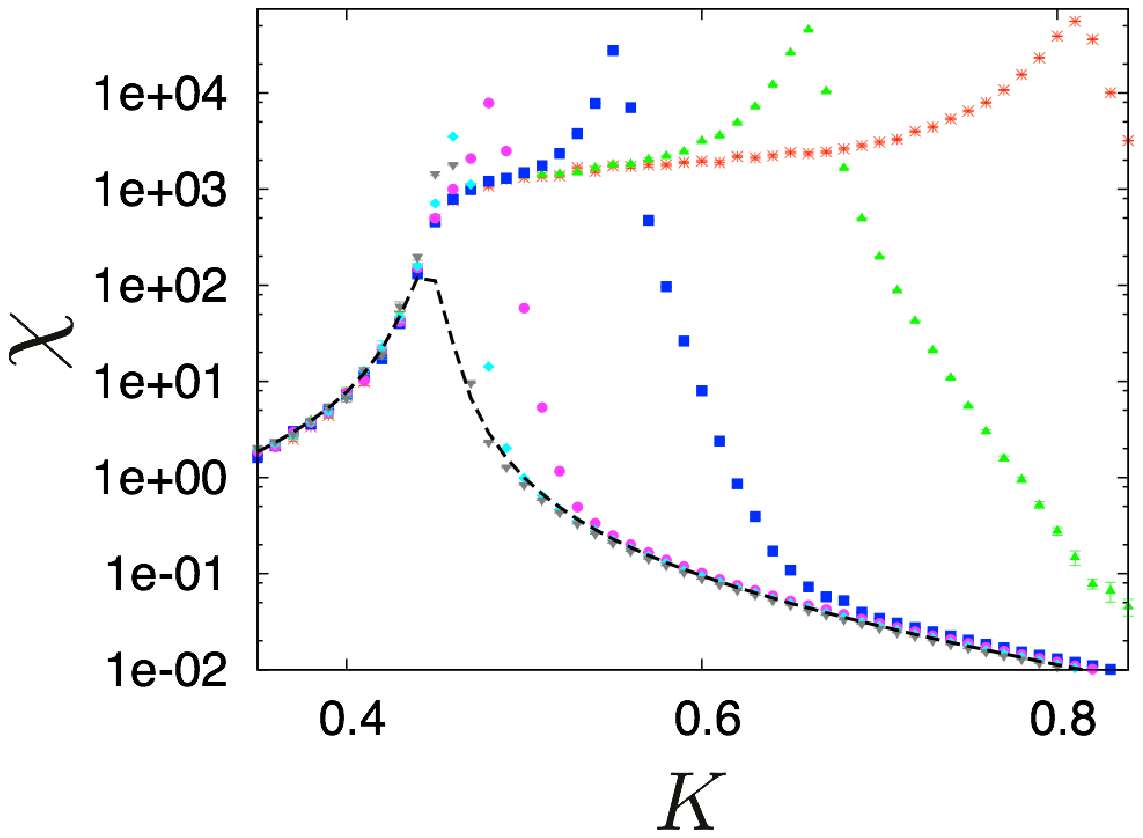}}
\caption{Thermodynamic quantities for the system shown in Fig.~\ref{fig:geom2d} with $\mathcal{N}=10$ and   $L_{0}=100$ connected (in a periodic way) by strips of size $100 \times M$ as a function of the coupling $K$: energy per spin (upper panel, left) $e$, heat capacity $C$ (upper panel, right), magnetization per spin $m$ (lower panel, left) and magnetic susceptibility $\chi$ (lower panel, right). Results for a single OBC  square $L_{0}=100$ are plotted with a black dashed line for comparison.}
\label{fig:2d_10-100}
\end{figure}

Now we consider a 2D array of a linear size $\mathcal{N}$ consisting of $\mathcal{N}^{2}$  squares of size $L_{0}$ connected by strips (channels) of the length $L$ and the width $M$ as shown in Fig.~\ref{fig:geom2d}, where we also  indicate the lines along which we compute the spin-spin correlation functions. 
In Fig.~\ref{fig:2d_10-100} we plot the thermodynamic quantities for this system as a function of $K$ for several values of the channel width $M$.
For the 2D array we observe a qualitatively similar scenario as for the 1D array, i.e., the behaviour of both the energy density and the heat capacity closely follows that for a single square, whereas the inflection point of the magnetization and the peak in the magnetic susceptibility are shifted to larger values of $K$; the length of this shift is controlled by the channel width $M$. A significative difference between 1D and 2D systems is that the latter exhibit a true ordering transition in the thermodynamic limit. This means that the maximum  in the susceptibility grows to infinity upon increasing the size of the 2D array, while for 1D arrays this is not the case. A shoulder in the magnetic susceptibility is a ghost of the rounded phase transition in the 2D square. 

\begin{figure}[h]
\includegraphics[width=0.49\textwidth]{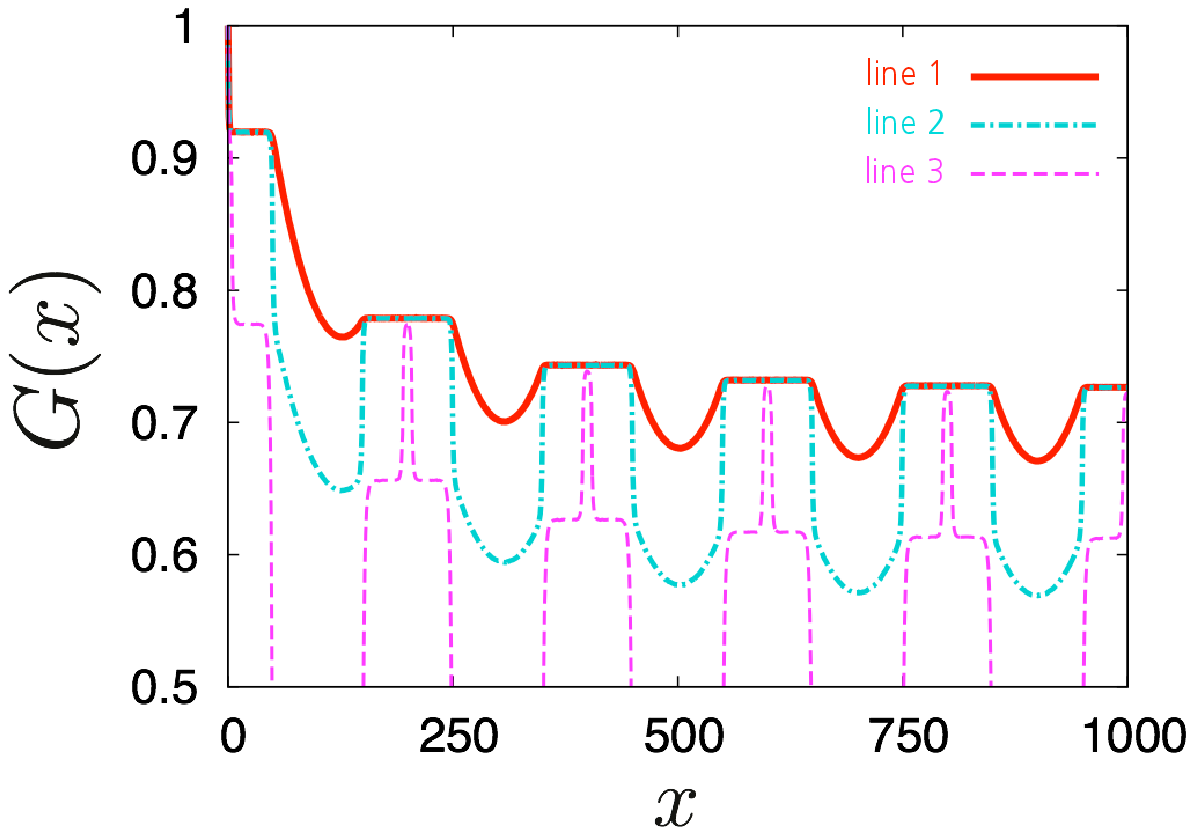}
\includegraphics[width=0.49\textwidth]{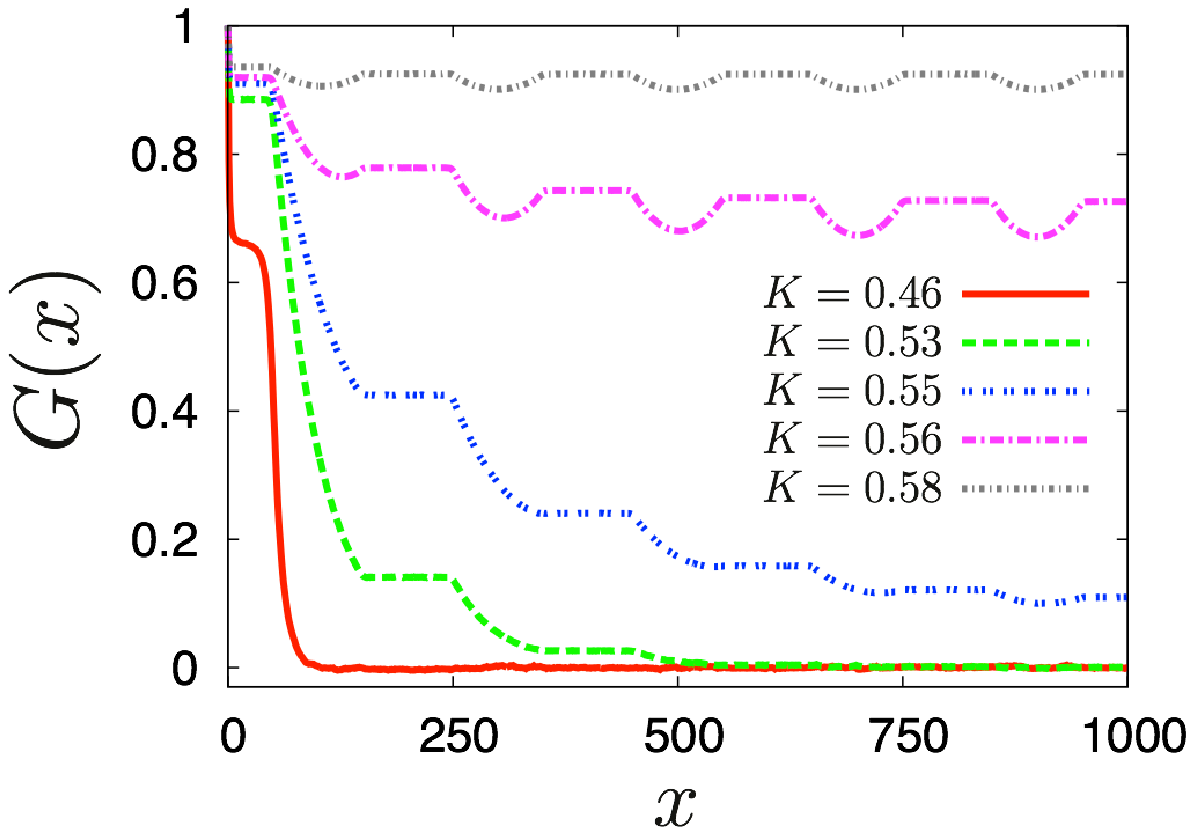}
\caption{Left panel: the spin-spin correlation function $G(x)$ as a function of the distance $x$ for a system with $L_{0}=100$, $L=100$, $M=10$ for $K =0.56$ along the three different lines: centers of the channels (red solid line), sides of the channels (cyan dash-dotted line) and along the sides of the squares (magenta dashed line). Right panel: the correlation function along the centers of the channels for various couplings, $K=0.46,0.53,0.55,0.56,0.58$.}
\label{fig:2d_10-100_corr}
\end{figure}

In Fig.~\ref{fig:2d_10-100_corr} we plot the spin-spin correlation function $G(x)$ for the 2D system. In this case the correlations along the side of the square are weaker than in the central part, but they reach the value of $G(x)$ in the central part at the  points of the  channel cross-section. We observe that below a certain value of  $K$ (roughly between 0.55 and 0.56)  $G(x)$ does not decay to zero. This is a clear indication of the existence of order in the network. The transition as signaled by the peak of the susceptibility which occurs at $K\approx 0.55$, in agreement with the Fisher-Privman theory.

%====================================
%================= SECTION 3.D
%====================================
\subsection{Computation of the effective coupling.} 
\begin{figure}[htbp]
\includegraphics[width=0.45\textwidth]{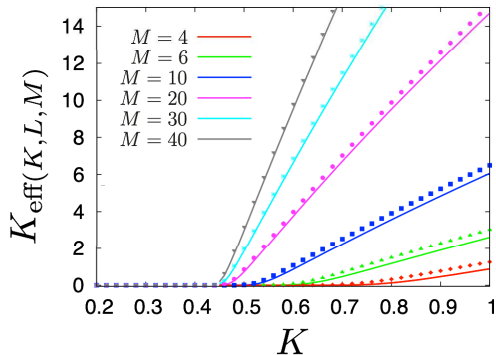}
\caption{The effective interaction constant $K_{\textrm{eff}}$ mediated by  the  Ising strip (channel) of the size $L \times M$ as a function of $K$ for $L=100$ and $M=4,6,10,20,30,40$. Lines correspond to eq.~(\ref{Boltzmann}), symbols correspond to the MC results eq.~(\ref{eq:Jeff_mc}).}
\label{fig:Jeff}
\end{figure}

We can compute the effective coupling constant between two spin boxes mediated by the channel  numerically. Let us consider the strip of the length $L$ and the  width $M$. Along two sides of length $L$ the OBC are applied, whereas two sides of the length $M$ are subjected to the surface fields $H_{1}^{-}$ (left side, $x=1$) and $H_{1}^{+}$ (right size, $x=L$). Further, we assume that the field at the right side  is $H_{1}^{+}=+1$ and consider two cases for  the field at the left side $H_{1}^{-}=-1,+1$. We denote the  free energy of the system for these cases $F_{++}$ and $F_{-+}$, respectively. The surface magnetization at  the left side of the strip is $M_{1}^{-}=\sum_{x=1} \sigma_{j}$ whereas at the right side of the strip is $M_{1}^{+}=\sum_{x=L} \sigma_{j}$. Using the temperature integration method, for $H_{1}^{+}=|H_{1}^{-}|=1$ we can compute free energies:
\be
\beta F_{++}(\beta,L,M)= \int \limits_{0}^{\beta}
\la E+M_{1}^{-}+M_{1}^{+} \ra_{\beta',L,M} {\rm d}\beta^{\prime} \, ,
\ee
and
\be
\beta F_{-+}(\beta,L,M)= \int \limits_{0}^{\beta}
\la E+M_{1}^{-}-M_{1}^{+} \ra_{\beta',L,M} {\rm d}\beta' \, ,
\ee
where the average $\la \dots \ra$ is performed for a given geometry $L \times M$ and inverse temperature $\beta'$. The effective interaction constant is thus given by 
\be
\label{eq:Jeff_mc}
K_{\textrm{eff}}(\beta,L,M)=\frac{1}{2}\beta\left(F_{-+}(\beta,L,M)-F_{++}(\beta,L,M) \right) \, .
\ee
Our prediction for the effective interaction constant based on the extended Fisher-Privman theory is given by eq.~(\ref{Boltzmann}). In Fig.~\ref{fig:Jeff} we plot MC results (eq.(\ref{eq:Jeff_mc})) for $K_{\textrm{eff}}(\beta,L,M)$ for $L=100$ and $M=4,6,10,20,30,40$ as functions of $K$ in comparison with the theoretical results predicted by (\ref{Boltzmann}). We observe an excellent agreement between the MC data and the aforementioned theoretical curve.

%====================================
%================= SECTION 3.E
%====================================
\subsection{2D arrays of cubes}
\begin{figure}[h]
\mbox{
\includegraphics[width=0.49\textwidth]{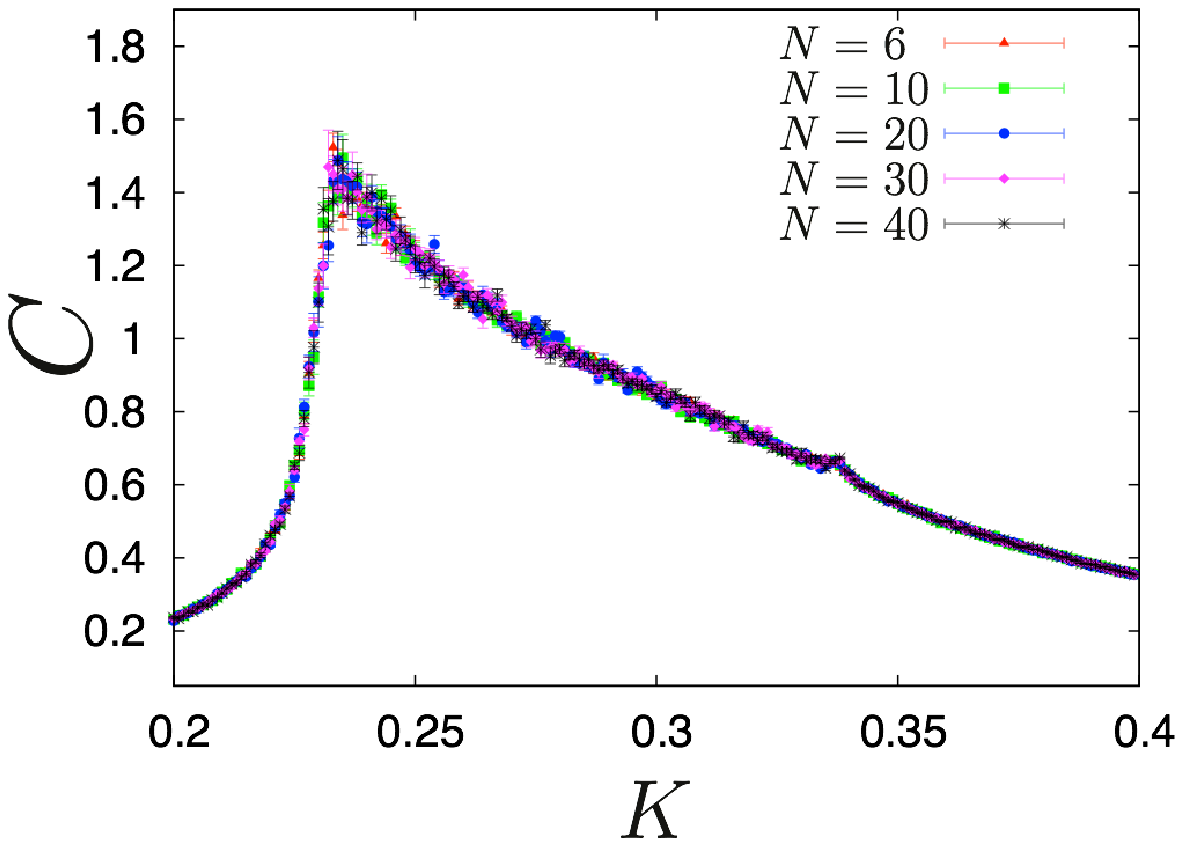}
\includegraphics[width=0.49\textwidth]{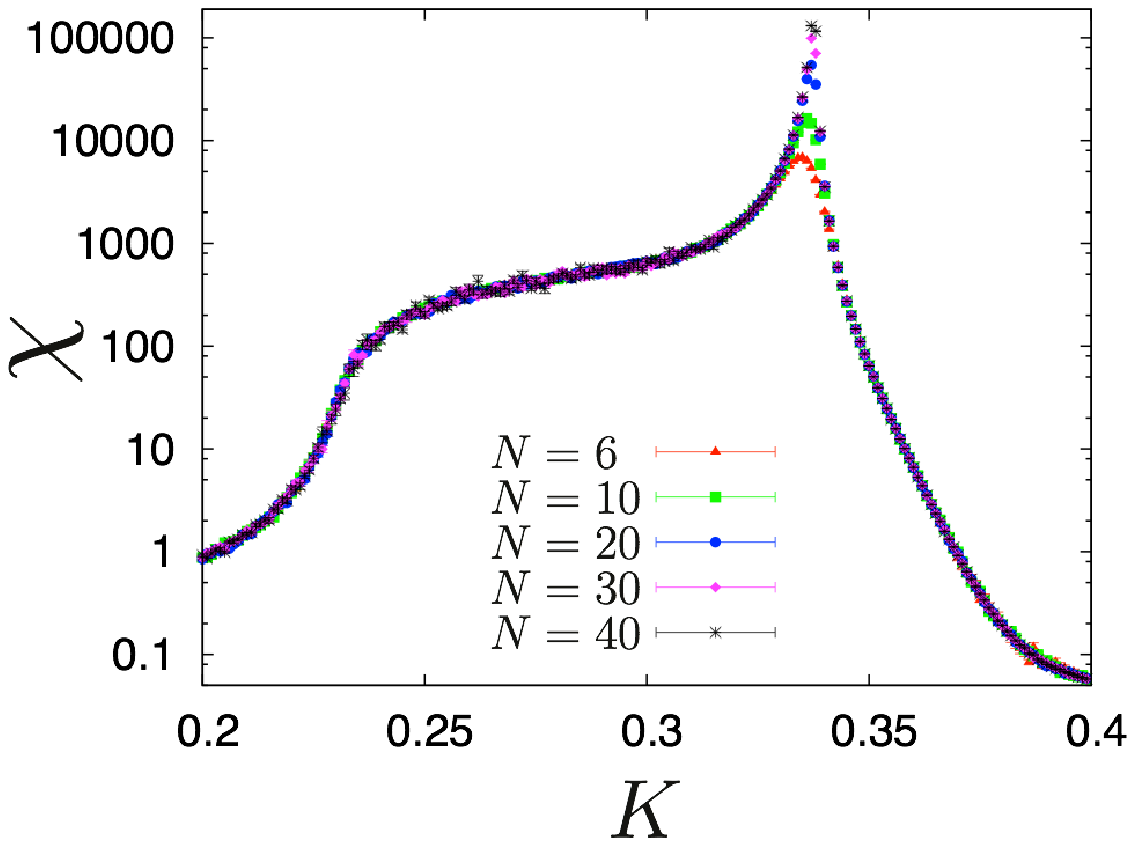}}
\caption{Thermodynamic quantities for the 2D array of $\mathcal{N}\times \mathcal{N}$ cubes of side $L_{0}=20$ connected (in a periodic way) by channels of size $40 \times 4 \times 4$ as functions of the coupling $K$: heat capacity $C$ (left panel), magnetic susceptibility $\chi$ (right panel).}
\label{fig:3d_N}
\end{figure}

In the same way we have performed various simulations of a 2D array of the size $10\times 10$ of 3D cubes of the size $L_{0}=40$, connected by channels of the size $L \times M \times M$ with $L=40$ and $M=2,4,6,10,16$. The geometry of the coarse-grained system is exactly the same, as for the 2D system of Fig.~\ref{fig:geom2d}, but now it consists of 3D cubes and is connected by 3D channels. As before, we have compared the results for various thermodynamic quantities with those  for a single OBC cube of the same size $L_{0}=40$. We have observed that various thermodynamic quantities behave in the same way as for the 2D system. In Fig.~\ref{fig:3d_N} we plot both the heat capacity and the susceptibility as functions of the coupling $K$ for $L_{0}=20$ and the channel size $40 \times 4 \times 4$ for various values of the number of cubes $\mathcal{N}$. As expected, the heat capacity does not change by increasing the number of cubes of the network. The magnetic susceptibility exhibits a maximum whose amplitude increases with $\mathcal{N}$, while its location remains almost unchanged. We note that contrary to the 2D systems, the heat capacity $C$ exhibits a small  peak at the value of $K$ which roughly corresponds to the maximum of the magnetic susceptibility $\chi$.

\begin{figure}[h]
\mbox{
\includegraphics[width=0.49\textwidth]{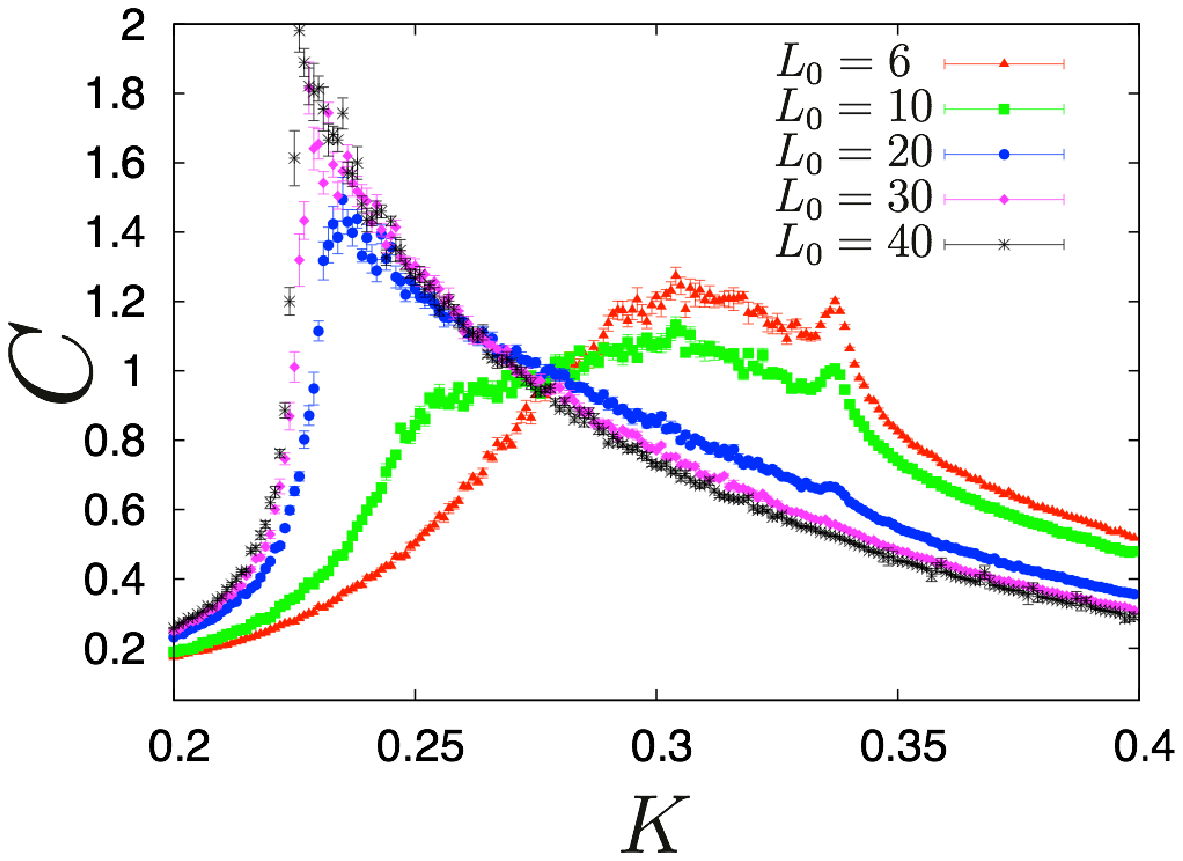}
\includegraphics[width=0.49\textwidth]{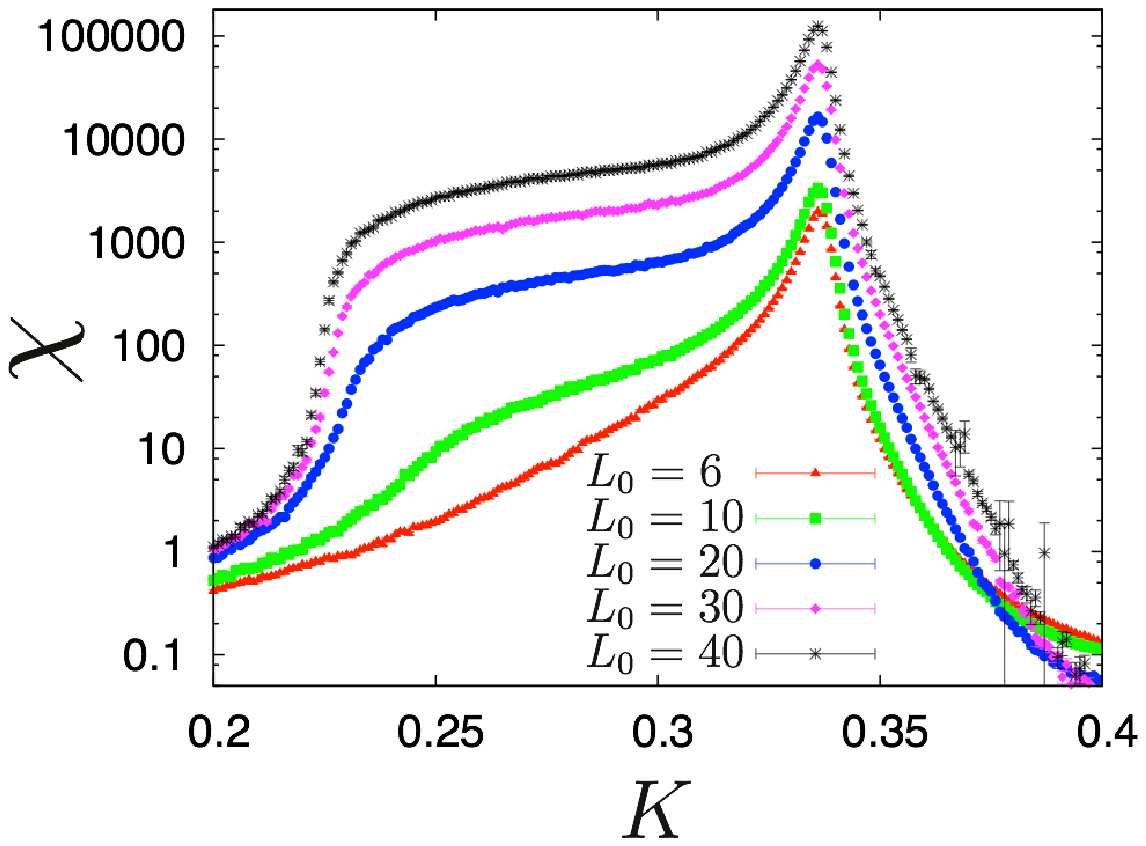}}
\caption{Thermodynamic quantities for the 2D array of $10\times 10$ cubes of the side $L_{0}$ connected (in a periodic way) by channels of size $40 \times 4 \times 4$ as functions of the coupling $K$: heat capacity $C$ (left panel), magnetic susceptibility $\chi$ (right panel).}
\label{fig:3d_L0}
\end{figure}

\begin{figure}[htbp]
\mbox{
\includegraphics[width=0.49\textwidth]{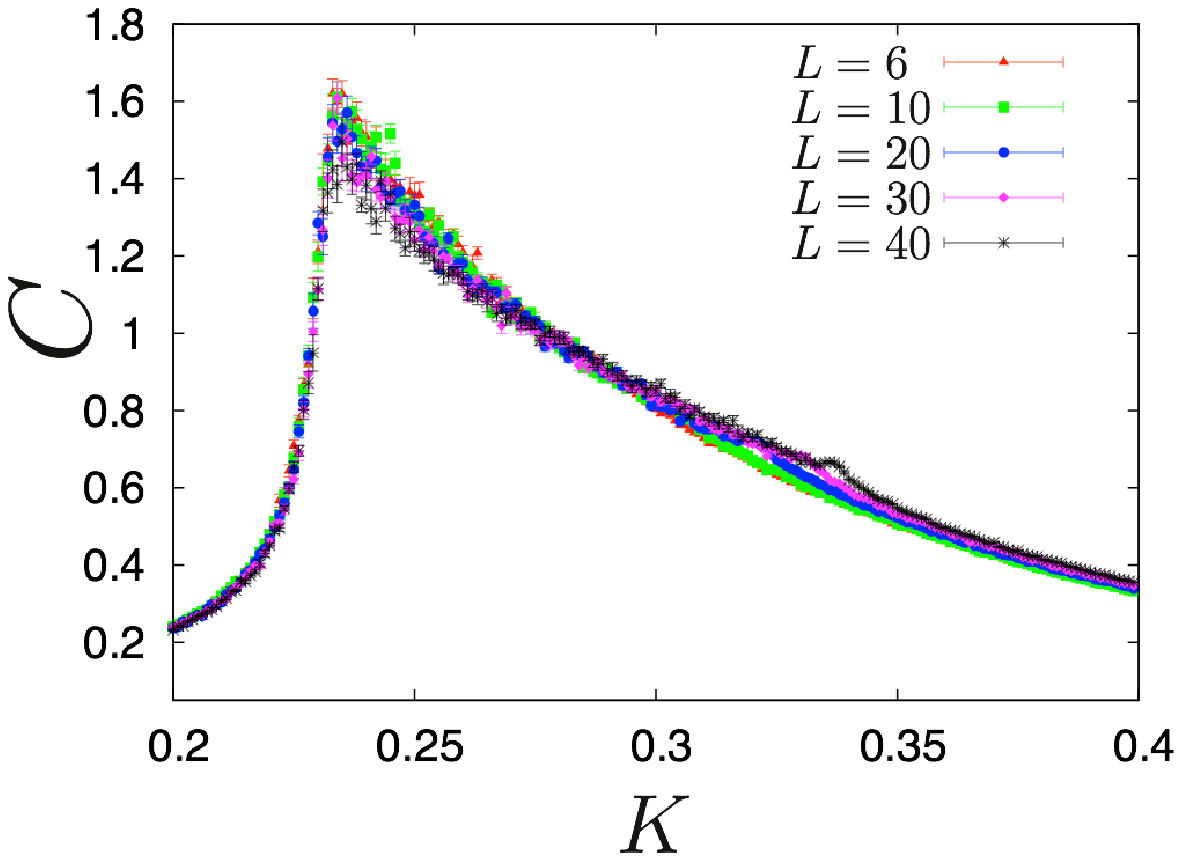}
\includegraphics[width=0.49\textwidth]{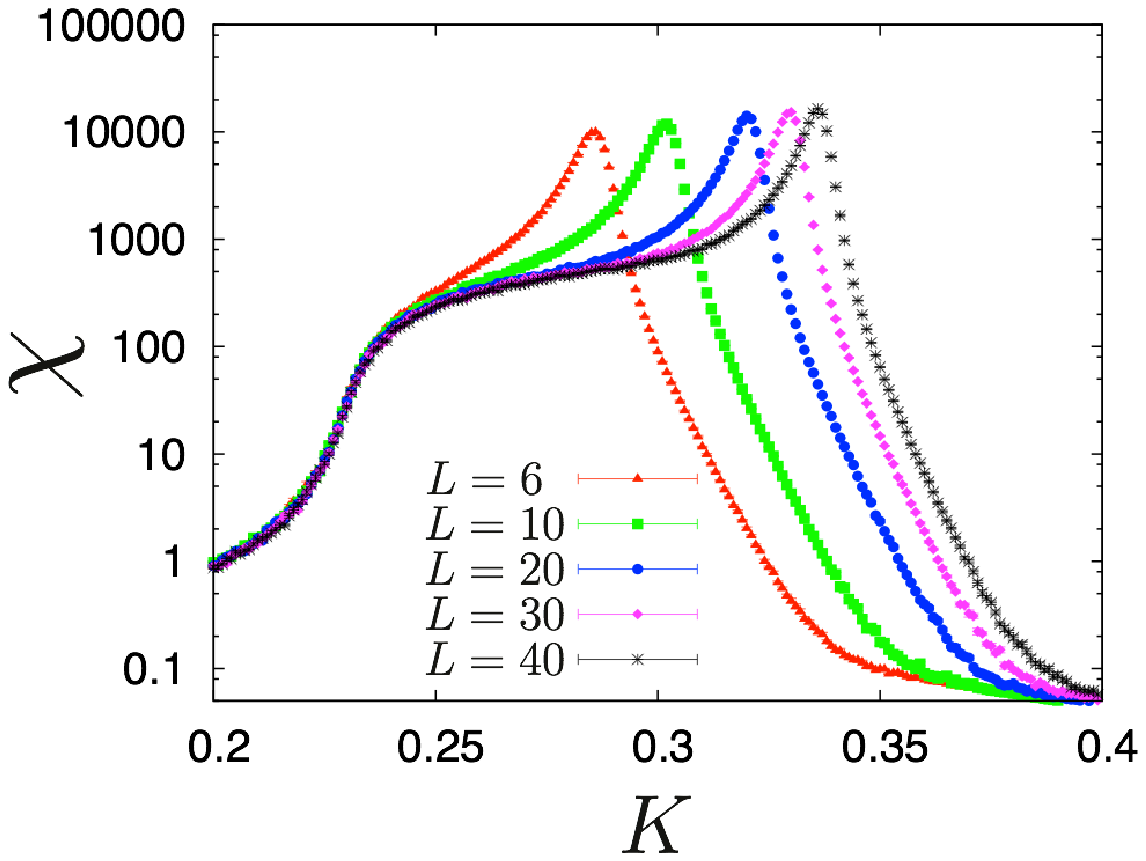}}
\caption{Thermodynamic quantities for the 2D array of $10\times 10$ cubes of the side $L_{0}=20$ connected (in a periodic way) by channels of the  size $L \times 4 \times 4$ as functions of the coupling $K$: heat capacity $C$ (left panel), magnetic susceptibility $\chi$ (right panel).}
\label{fig:3d_L}
\end{figure}

\begin{figure}[h]
\mbox{
\includegraphics[width=0.49\textwidth]{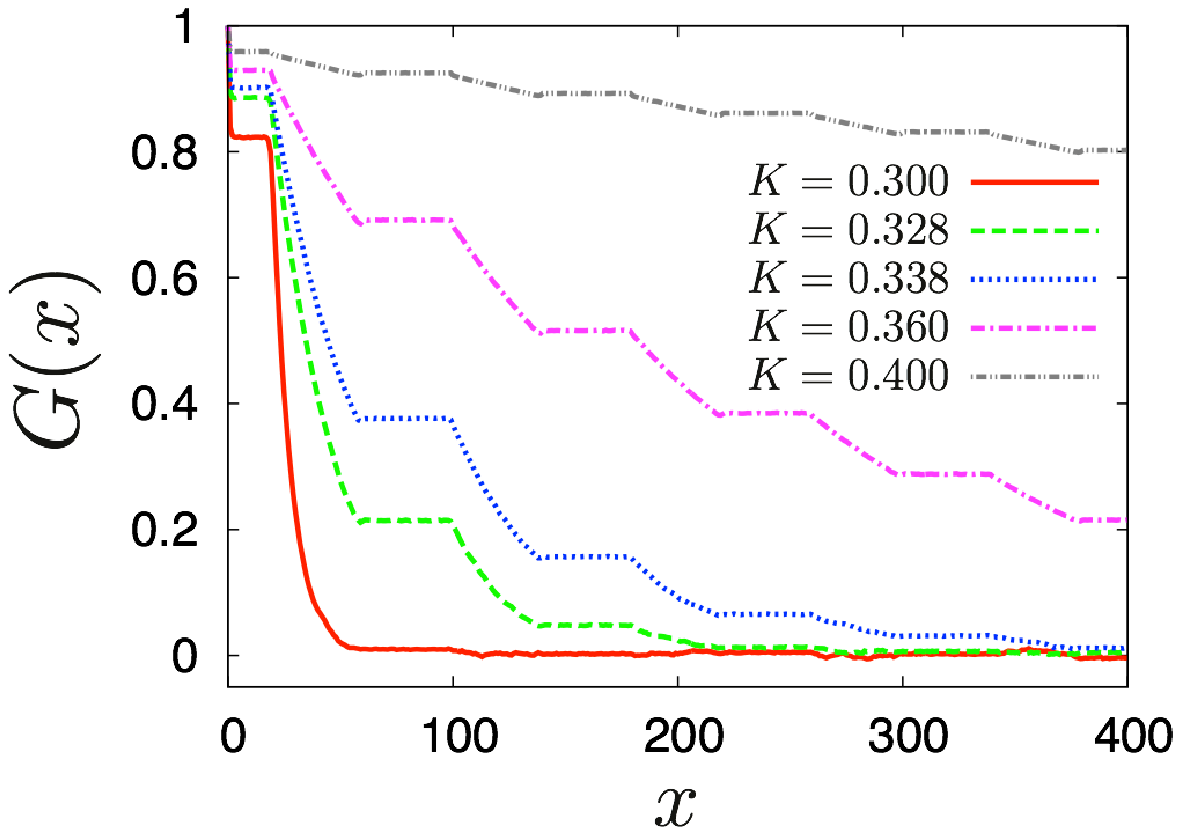}
\includegraphics[width=0.49\textwidth]{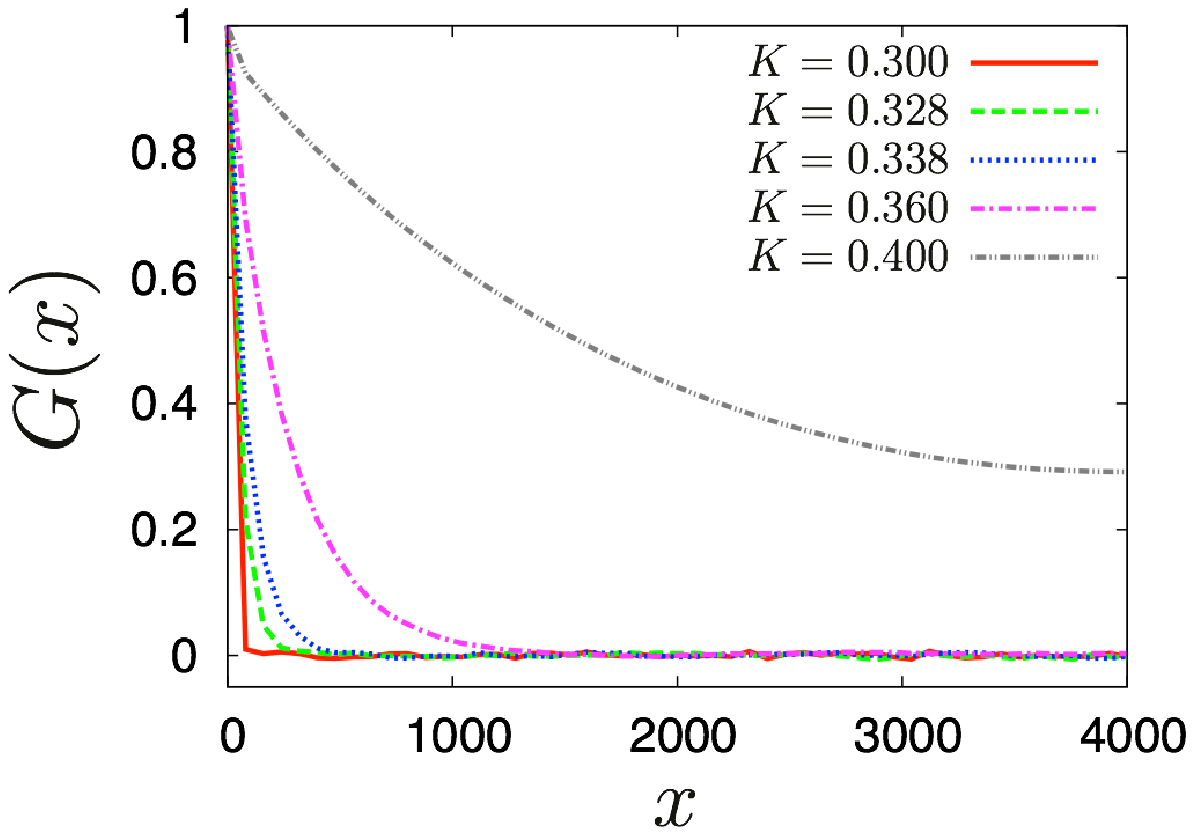}}
\caption{Left panel: the spin-spin correlation function $G(x)$ as a function of the distance $x$ along the center of the channel, (line 1 of Fig.~\ref{fig:geom2d}(a)), for the 1D array of  $\mathcal{N}=100$ 3D cubes of the linear size $L_{0}=40$ (periodically) connected by channels of the size $40 \times 4 \times 4$ for several values of the coupling $K$. Right panel: the corresponding values of $G(x)$ at the centers of the cubes.}
\label{fig:3d_corr}
\end{figure}

In Fig.~\ref{fig:3d_L0} we plot the results for the 2D array of $10\times 10$ cubes connected by channels of the size $40 \times 4 \times 4$ for various cube sizes $L_{0}$. For small cubes, $L_{0} \le 10$, the heat capacity  forms  a wide graph with two maxima, a sharp one at  $K \simeq 0.34$ (which coincides with the maximum of susceptibility) and the second, broad one. As we increase $L_{0}$, the two maxima merge into a single one which gradually increases and shifts toward  $K_{c}\simeq 0.2216$, the critical coupling of the 3D Ising model. The susceptibility has a single pronounced maximum at $K \simeq 0.34$. As we increase $L_{0}$, the plateau (shoulder) between this point and $K_{c}$ is formed.
In Fig.~\ref{fig:3d_L} we plot the thermodynamic quantities for the 2D array $10\times 10$ of cubes of the side $L_{0}=20$ for various channel lengths and fixed cross-section equal to $4\times 4$. We observe that the small maximum of the heat capacity changes its position with $L$ in consistence with the behaviour of the susceptibility maxima, the latter move toward smaller values of $K$ as the channel length $L$ is increased.
Finally, we have computed the spin-spin correlation function $G(x)$ for the $1D$ array of 3D cubes as a function of the distance $x$ along the center of the channel (red solid line in Fig.~\ref{fig:1d_100-100_corr1}(a)). Comparing Figs.~\ref{fig:1d_100-100_corr2} and \ref{fig:3d_corr} we can conjecture that the function $G(x)$ in the 1D array of cubes behaves in the same way as $G(x)$ for the 1D array of squares. The only difference is the range of couplings $K$ for which the correlations spread across the whole system - for the 1D array of 3D cubes this occurs at much smaller values of $K$ (larger temperatures).

%====================================
%================= CONCLUSIONS
%====================================
\section{Conclusion}
In this paper we have presented in  details the theory and  the MC simulations which explain how an Ising-like system forming a 2D array of  boxes connected by narrow channels can support a long range order on length scales much larger than the bulk correlation length. We show that for a given temperature and width of the 2D channel  there exists a critical length of the latter such that the network of boxes  is ordered when the channels does not exceed that critical length. Such a theoretical analysis follows from an effective temperature-dependent coupling constant between the boxes that we determined analytically and tested against numerical simulations. Eventually we have extracted the phase diagram of the planar network of 2D systems. The observed cooperative effect follows from the existence of an emerging length scales that develops inside the connecting channels and dominates over length scales much larger than the ones of bulk fluctuations. The Fisher-Privman theory plays a crucial role in our thinking; for 2D systems we show how important point tension (the analogue of the line tension in two dimensions) is in considerations of the validity of this theory, which we subject to a test using the exactly-solvable theory of Ising strips. For the planar network of 3D boxes connected by rods we have provided only the MC simulation results. The extension of the Fisher-Privman theory to this case is a subject of our future work. The cooperative phenomenon that we have found in our system is analogous to the one observed experimentally in superfluid $^4$He \cite{PKMG} and is a  consequence of phase transitions and critical phenomena in confined geometries. The mechanism for the emerging action-at-a-distance which we have described should work for classical binary liquid mixtures at two-phase bulk coexistence - provided that the surfaces of cells and channels have no preference for any of the two phases.

%====================================
%================= ACKNOWLEDGMENTS
%====================================
\subsection*{Acknowledgments}
D. B. A. acknowledges the kind hospitality of Prof S. Dietrich and the Max Planck Society for multiple visits while this work was done. A.S. wishes to acknowledge the warm hospitality of the Rudolf Peierls Center for Theoretical Physics (Oxford) where part of this work were done.

\end{document}